\documentclass[11pt,a4paper]{elsarticle}

\usepackage[utf8]{inputenc}
\usepackage[UKenglish]{babel}
\usepackage{natbib}
\usepackage{comment}
\usepackage{setspace}

\usepackage[left=1.75cm,right=1.75cm,top=1.75cm,bottom=1.75cm]{geometry}

\usepackage{url}        %
\usepackage{amssymb}    
\usepackage{lscape}     
\usepackage{booktabs}   
\usepackage{amsmath}
\usepackage{caption}
\usepackage{subcaption}
\usepackage[]{svg} 
\usepackage[per-mode=symbol]{siunitx}
\usepackage{eurosym} 
\usepackage{placeins} 
\usepackage{float}
\usepackage{hyperref}
\usepackage{algorithm}
\usepackage{algpseudocode}
\usepackage[makeroom]{cancel}

\usepackage{tikz}
\usetikzlibrary{shapes.geometric, arrows}

\DeclareSIUnit{\eur}{\text{\euro}}
\DeclareSIUnit{\cotwo}{CO\ensuremath{_2}}
\DeclareSIUnit{\year}{yr}
\DeclareSIUnit{\MWh}{MWh}
\DeclareSIUnit{\kWh}{kWh}
\DeclareSIUnit{\ct}{ct}

\usepackage{multirow}
\usepackage{makecell}

\usepackage{graphicx}
\usepackage[export]{adjustbox}

\usepackage{tabularx}
\usepackage{array}
\usepackage{tabu}
\usepackage{multirow}
\usepackage[dvipsnames]{xcolor}
\usepackage[table]{xcolor}   
\usepackage{booktabs}        
\usepackage{geometry}        
\usepackage{pdflscape}          
\usepackage{array}              
\usepackage{makecell}           
\usepackage{wrapfig}
\usepackage{graphicx}

\begin{document}

\onehalfspacing

\begin{frontmatter}

\title{Value-focused modelling to generate alternatives -- Coupling multi-criteria decision analysis and optimisation models to support strategic decisions}

\author[1]{Emily Bergup}
\ead{emily.bergup@rub.de}

\author[1]{Jonas Finke\corref{cor1}}
\ead{jonas.finke@rub.de}

\author[2]{Sebastian Schär}
\ead{sebastian.schaer@eawag.ch}

\author[1]{Valentin Bertsch}
\ead{valentin.bertsch@rub.de}

\cortext[cor1]{Corresponding author}

\address[1]{Ruhr-Universität Bochum, Universitätsstr.\ 150, 44801 Bochum, Germany}

\address[2]{Swiss Federal Institute of Aquatic Science and Technology (Eawag), Department Environmental Social Sciences, Überlandstrasse 133, 8600 Dübendorf, Switzerland}

\begin{abstract}
Decision support methods from operations research are widely used to support complex planning decisions. Within the energy sector, energy system models (ESMs) applying modelling to generate alternatives (MGA) generate large sets of near-optimal but very different system configurations. However, they typically generate and analyse alternatives in the model variable space without ensuring stakeholder relevance. 
Multi-criteria decision analysis (MCDA), in contrast, provides a structured means to account for conflicting objectives and heterogeneous stakeholder interests but often relies on a limited set of pre-defined alternatives that may not appropriately represent the feasible solution space. To address these limitations, this work proposes \textit{value-focused modelling to generate alternatives (VF-MGA)}, a novel methodology that bidirectionally couples MGA and MCDA.
Stakeholder objectives elicited within the MCDA inform the MGA-algorithm, enabling a stakeholder-orientated diversification of the alternatives, which are subsequently evaluated within the MCDA based on elicited stakeholder preferences, thereby providing a comprehensive decision basis.
Applied to a case study on the decarbonised energy supply of a large university campus, involving eleven stakeholders representing diverse institutional groups, VF-MGA (i) systematically integrates stakeholder objectives into the generation of 691 alternatives reflecting stakeholder-relevant interests, (ii) enables the identification of stakeholder-relevant alternatives from this large set through MCDA-based evaluation, and (iii) provides more differentiated stakeholder preference information by evaluating a large and diverse set of alternatives, thereby revealing acceptable ranges for system options. With this, VF-MGA provides a generalisable methodology for complex planning decision integrating quantitative modelling with participatory decision analysis.
\end{abstract}

\begin{keyword}
Energy system model \sep multicriteria decision analysis \sep Modelling to generate alternatives \sep Value-focused thinking
\end{keyword}

\begin{highlights} 
 \item Integrate multi-criteria decision analysis and near-optimal optimisation modelling 
 \item Stakeholder objectives guide model-based generation of near-optimal alternatives
 \item Elicitation of objectives and preferences with structured stakeholder interviews
 \item MCDA evaluates 691 energy transition alternatives based on stakeholder preferences
 \item MCDA ranking simplifies solution space to highlight conflicts/consensus
\end{highlights}

\end{frontmatter}


\section{Introduction}

Complex planning decisions are characterised by a large decision space, multiple, often conflicting objectives, and the need to account for diverse stakeholder perspectives. Identifying and evaluating relevant decision alternatives in such settings is therefore a challenging task for which guidance is needed. 
These challenges are particularly pronounced in the context of energy system transformation, a domain that exemplifies complex planning decisions due to its high system complexity and diverse stakeholder interests. 
Energy system models (ESMs) and multi-criteria decision analysis (MCDA) are two distinct approaches that can offer the necessary guidance in different ways \citep{Prina.2020, Greco.2016, Cinelli.2020}. ESMs, on the one hand, are computational models that aim to represent the interactions, dynamics, and interdependencies between various components of energy systems \citep{Pfenninger.2014, Kondili.2010}. By exploring technically feasible system configurations, they can generate and analyse potential transition pathways \citep{Prina.2020}. MCDA, on the other hand, provides a structured decision-support framework that evaluates and compares alternatives with respect to multiple objectives while explicitly incorporating stakeholder preferences \citep{Belton.2002}. MCDA thereby supports decision-making by structuring trade-offs and guiding the selection among competing alternatives \citep{Diakoulaki2005}. In the following, we discuss key limitations of both approaches and outline how this study aims to address them.

\subsection{Relevant literature and research gaps}
Modelling to generate alternatives (MGA) is a well-established approach to overcome several limitations of traditional optimisation-based modelling and has recently been increasingly applied to ESMs \citep{DeCarolis.2011, DeCarolis.2016, Yue.2018, Lau.2024}. MGA refers to a set of different optimisation techniques that aim to generate structurally diverse solutions within the near-optimal solution space \citep{Brill.1982}. In the context of ESMs, structural diversity often refers to different system configurations, including alternative technology portfolios and the spatial allocation of installed capacity \citep{Lombardi.2020, Finke.2026}. By exploring the near-optimal solution space rather than focusing on single optimal solutions, MGA allows for a broader range of technically feasible system configurations to be analysed \citep{Trutnevyte.2016}.
MGA can also address both parametric and structural uncertainty inherent to energy system modelling \citep{Yue.2018}. Parametric uncertainty concerns the uncertainty of input data or model parameters based on incomplete information and unpredictable future developments. Structural uncertainty emerges from the fundamental challenge of abstracting complex real-world problems into simplified model structures that are not able to fully capture the real-world complexity \citep{Lau.2024, Price.2017}. With regard to structural uncertainty, MGA assumes that the real-world decision space is defined by technical system requirements and a maximum acceptable cost level \citep{Lau.2024}. Exploring this near-optimal solution space enhances the likelihood that objectives not explicitly represented in the model are indirectly reflected, rather than limiting the analysis to a single optimal solution or the non-inferior frontier \citep{DeCarolis.2011}. With regard to parametric uncertainty, diverse, near-optimal model outcomes may approximate parameter perturbations in the objective function coefficients \citep{DeCarolis.2011}. By additionally investigating the intersection of near-optimal solution spaces resulting from different scenarios or parameter assumptions, the identification of alternatives that remain robust under varying assumptions is enabled \citep{Grochowicz.2023}. 
Despite these advantages, two important shortcomings remain when using ESMs applying MGA (ESM-MGA) to support real-world decision-making. In the following, we refer to these shortcomings as \textit{(MGA Gap 1)} and \textit{(MGA Gap 2)}. 

\textit{(MGA Gap 1)} Exploring the near-optimal solution space does not explicitly ensure that the resulting alternatives adequately reflect stakeholder interests \citep{DeCarolis.2011}.
As MGA is typically applied without explicit stakeholder input, diversity emerges from a set of \emph{modelling} decisions that determine how and along which system dimensions alternatives are diversified \citep{Lombardi.19.07.2024}. As a result, while technically feasible alternatives are identified, their alignment with stakeholder interests in relation to the underlying decision problem is not explicitly ensured and may therefore arise only incidentally \citep{Lombardi.2023, Lombardi.19.07.2024}. Participatory modelling aims to address this gap by incorporating stakeholder perspectives into the modelling process. However, the majority of studies applying ESMs lack well-structured stakeholder interactions \citep{McGookin.2024}.

\textit{(MGA Gap 2)} Additionally, the systematic diversification of alternatives typically results in very large sets of feasible system configurations, which can be overwhelming to assess \citep{Esser.2024}. Existing MGA analyses commonly examine patterns in the system structure within the set of alternatives, for instance, by identifying must-have, real-choice, or must-avoid technologies (e.g. \citet{DeCarolis.2016, Pickering.2022}), by investigating structurally different system configurations (e.g. \citet{DeCarolis.2011, Esser.2024, Berntsen.2017}), or by examining correlations between technologies (e.g. \citep{Neumann.2021, Finke.2024}). However, these analyses remain largely confined to the variable decision space defined by the model. They do not evaluate the resulting alternatives in the same objective space as a broad set of societal stakeholders. Instead, model results typically describe technical system configurations, whereas stakeholders assess alternatives on a broader set of objectives, some of which are not directly represented in the model outputs \citep{McGookin.2021}. Furthermore, they may have substantially different trade-off preferences between objectives compared to those underlying the model.
Consequently, in the absence of a structured evaluation framework and limited dimensionality, ESM-MGA provide limited guidance for decision makers on how to interpret the generated alternatives and identify  feasible solutions across heterogeneous stakeholder interests. 

MCDA methods offer a promising opportunity to address these remaining limitations in a targeted manner and are an established decision-support framework in the context of energy systems \citep{Wang.2009, Cinelli.2022}. They offer a structured approach for evaluating different decision alternatives, considering diverse, often conflicting objectives while explicitly incorporating heterogeneous stakeholder preferences \citep{Belton.2002}. MCDA enables the integration of both quantitative and qualitative objectives, measured on different scales, thereby supporting the identification of best-performing alternatives related to stakeholder-relevant objectives and preferences \citep{Kiker.2005}.
Despite these advantages, current applications of MCDA in the energy sector also exhibit two crucial limitations, labelled as \textit{(MCDA Gap 1)} and \textit{(MCDA Gap 2)}.

\textit{(MCDA Gap 1)} The effectiveness of MCDA highly depends on the quality and relevance of the alternatives selected for evaluation \citep{Hamalainen.2024}. The alternatives are often pre-defined by the decision-makers themselves, by researchers, or sourced from literature \citep{Rigo.2020}. More structured approaches to generate alternatives typically rely on problem-structuring methods from soft operational research rather than on quantitative models \citep{Franco.2022, Marttunen.2017}. However, their application becomes increasingly challenging when addressing complex system configurations with a large solution space.
Since MCDA evaluates a predefined set of alternatives, the analysis is inherently constrained by the alternatives considered. If this set does not adequately represent the feasible and desirable solution space, the alternatives identified as best-performing may only be preferred within a limited and potentially incomplete set.

\textit{(MCDA Gap 2)} Furthermore, many MCDA applications within the energy system field and broader operations research literature lack interaction with a broader range of relevant real-world stakeholders. 
Instead of eliciting preferences from a diverse set of stakeholders, studies frequently rely on simplifying assumptions, representative parameters, or input from a limited set of experts, often from academia \citep{Lienert.2026}.
This may lead to MCDA models that do not adequately reflect stakeholder preferences and, consequently, to results that lack practical validation and practical decision relevance as stakeholder preferences can have a considerable effect on the results of the MCDA \citep{Gregory.2020, Reichert.2019}.

To the best of our knowledge, no study has holistically addressed all of the aforementioned shortcomings associated with either ESM-MGA or MCDA. Nevertheless, several studies have made attempts to address individual aspects of the mentioned limitations. 

Within the MGA literature, \citet{Lombardi.19.07.2024} develop an approach to integrate stakeholder preferences into an MGA-based workflow by asking stakeholders to select preferred system configurations from an existing set of alternatives. However, no real stakeholders are involved, and the approach primarily reshapes the solution space in alignment with stakeholder preferences without enabling a systematic evaluation of alternatives with respect to stakeholder-relevant objectives. 
\citet{Vagero.2025} develop an interactive MGA-based tool that allows stakeholders to explore a large set of near-optimal energy system configurations. Although real stakeholders participate and the evaluation of ESM-MGA results is transferred to an objective space, the performance metrics are predefined by the authors rather than elicited from stakeholders. Furthermore, stakeholder preferences are primarily inferred indirectly from selected technology configurations and complemented by prioritisation of the predefined performance metrics, rather than being derived through structured preference elicitation.
\citet{Esser.2024} develop and apply a participatory modelling approach involving real stakeholders and formulate guidelines for transferring participatory MGA to other systems. In their approach, stakeholder participation mainly aims to ensure that the model adequately represents the real-world system and reflects the decision-making context relevant to stakeholders.
However, the diversification methods used to generate alternatives are not guided by stakeholder interests, and the analysis of the ESM-MGA results remains primarily in the variable space and does not translate the alternatives into a stakeholder-relevant objective space. The authors themselves note that stakeholders found the large set of generated alternatives difficult to interpret.
\citet{Sasse.2020} analyse electricity sector alternatives in Central Europe generated with ESM-MGA in relation to six objectives potentially relevant for electricity system transition. While this approach transfers the ESM-MGA solution space into an objective space, the objectives are defined by the authors rather than elicited from stakeholders, and the diversification of alternatives is not guided by stakeholder interests. Moreover, real stakeholders are not involved in the process.

Other studies attempt to address the identified limitations by coupling ESMs (without MGA) with MCDA. The vast majority of these approaches couple ESMs with MCDA one-directionally by evaluating model outputs ex post within an MCDA framework (e.g. \citet{Simoes.2019, Witt.2020, Choi.2020, Lerede.2021}). Such one-directional approaches leave the generation of alternatives unaffected by stakeholder preferences and therefore risk exploring solution spaces that may be of limited relevance to stakeholders. More generally, the alternatives evaluated in these studies represent only a limited subset of the feasible solution space, as conventional ESMs typically rely on a smaller number of predefined alternatives rather than systematically exploring near-optimal alternatives through diversification techniques such as MGA. 
We are only aware of one study from \citet{McKenna.2018} coupling an ESM and MCDA bidirectionally, by iteratively refining alternatives generated using an ESM based on discussions with real stakeholders to tailor decision alternatives more closely to stakeholders preferences. However, the study likewise does not use a formal MGA approach; hence, the diversification is limited, which means that the considered alternatives likely do not cover the complete decision space very well. The modifications to an initial set of alternatives were rather carried out ad-hoc following discussions with stakeholders, which limits reproducibility.

\subsection{Contributions}
To address the aforementioned limitations, we introduce \textit{value-focused modelling to generate alternatives (VF-MGA)}, a novel methodology that bidirectionally couples MGA and MCDA. Within the MCDA process, objectives are elicited in stakeholder interviews to inform and guide the MGA algorithm towards generating stakeholder-relevant alternatives. These alternatives are subsequently evaluated within the MCDA based on elicited stakeholder
preferences. VF-MGA follows the principles of value-focused thinking (VFT) \citep{Keeney.1992}, in which the fundamental values and objectives of stakeholders guide the analysis. VF-MGA is especially suited to the context of energy system modelling where
dedicated ESMs are employed. But it can also be applied in decision contexts with similar characteristics, e.g., facility location, infrastructure, or environmental management problems. 
As key contributions, this work:
\begin{itemize}
    \item[(i)]  conducts an MCDA study based on identifying a broad set of decision-relevant stakeholders and systematically eliciting their preferences, thereby addressing \textit{MCDA Gap 2};
    \item[(ii)] develops an approach through which stakeholder objectives derived from the MCDA inform and shape the generation of alternatives within the MGA framework, thereby addressing \textit{MGA Gap 1};
    \item[(iii)] evaluates a large set of MGA alternatives within the MCDA, thereby transferring the model's variable solution space into a stakeholder-relevant objective space, thereby addressing \textit{MGA Gap 2};
    \item[(iv)] uses this large and diverse set of model-derived decision alternatives to expand the decision space for MCDA, thereby addressing \textit{MCDA Gap 1};
\end{itemize}
Together, VF-MGA systematically integrates stakeholder preferences into the exploration and evaluation of a diverse set of model-based alternatives, thereby enabling more holistic decision support.

We apply the proposed VF-MGA methodology to a university campus energy system in order to demonstrate both its practical feasibility and to generate decision-relevant insights for campus energy planning. We use the Ruhr University Bochum (RUB), one of the largest universities in Germany, as a case study. With an annual energy demand of about 210 GWh, comparable to 12,000 German households, the decarbonisation of its energy supply is a major challenge. On the one hand, there is a wide range of feasible technical configurations to meet this demand; on the other hand, a diverse range of stakeholders and institutions is involved in or affected by the decision-making process. All these characteristics make it particularly suitable for demonstrating the VF-MGA methodology and deriving generalisable and transferable insights for similar decision contexts. 

The remainder of this paper is structured as follows. Section \ref{chapter: methods and data} introduces the VF-MGA methodology in more detail and outlines the underlying methodological foundations. Section \ref{chapter: case study} presents the application of VF-MGA in a case study of a large university campus energy system. Section \ref{chapter: results and discussion} reports and discusses the results of the analysis, including limitations and directions for future research. Finally, Section \ref{chapter: Conclusions and Outlook} provides a brief conclusion.

\section{Value-focused modelling to generate alternatives (VF-MGA)}
\label{chapter: methods and data}
\begin{figure}[!htb]
    \centering
    \includegraphics[width = 1\textwidth, trim= {2cm 2cm 1cm 0cm}, clip]{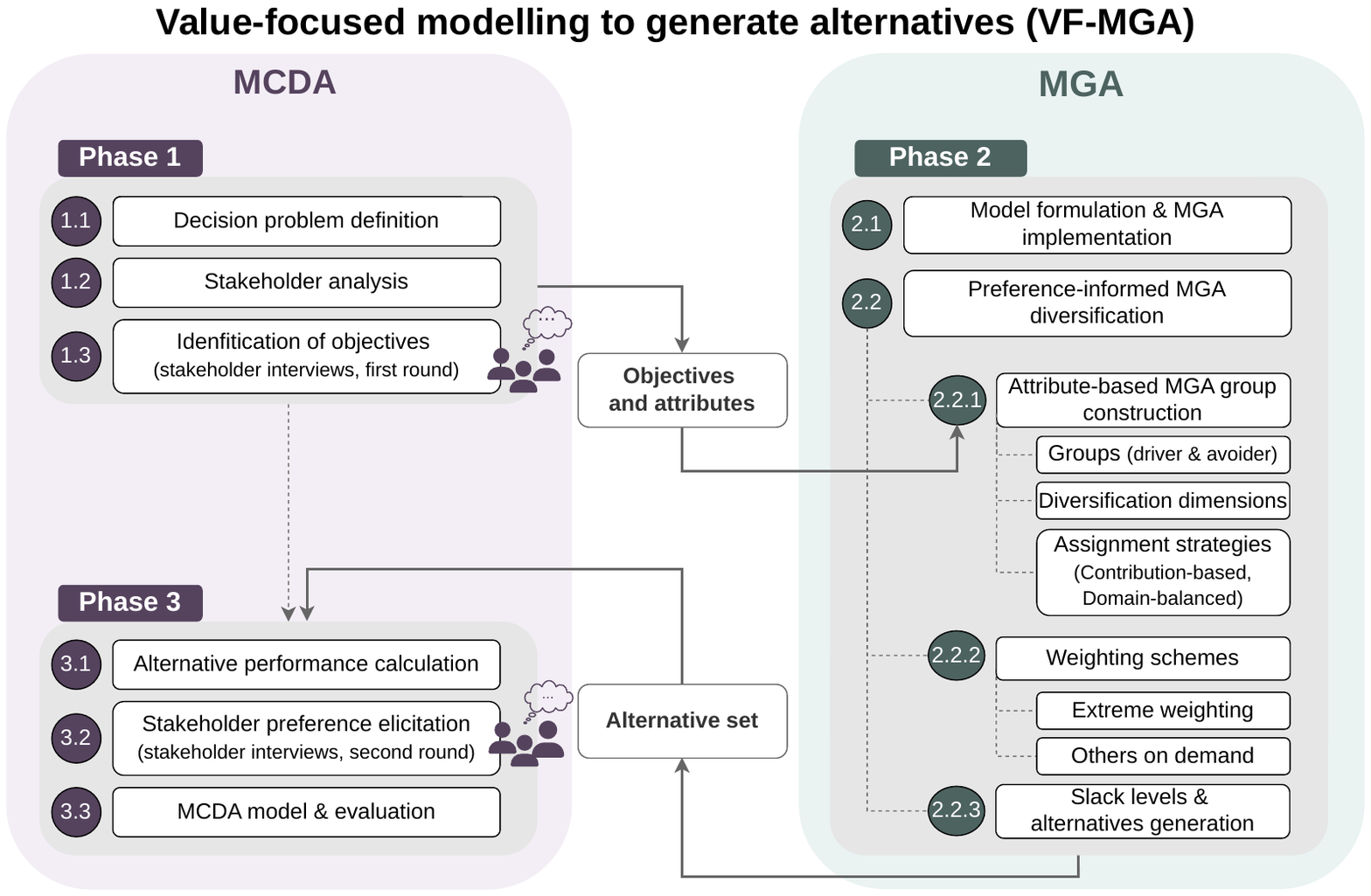}
    \caption{Methodology of value-focused modelling to generate alternatives (VF-MGA). VF-MGA couples MGA with MCDA in bidirectional manner and systematically exchanges information between both approaches. The numbering refers to the sequence of methodological steps in each of the three phases, while allowing for iterative refinement where appropriate. Systematic stakeholder interaction in steps 1.3 and 3.2 is crucial for the effectiveness of VF-MGA.}
    \label{fig:method_overview}
\end{figure}

The idea of VF-MGA is to combine and systematically exchange information between an optimisation model with MGA and an MCDA model to obtain a comprehensive set of alternatives which are informed by and can be evaluated on quantified stakeholder preferences. 
Generally, VF-MGA can be applied in various contexts, e.g., for infrastructure or environmental planning problems.
It requires (i) an optimisation model capable of applying MGA, i.e., the decision variables are suitable to create a broad set of meaningful alternatives within the decision context, and (ii) an MCDA method that provides the mathematical formalism for evaluating a large set of MGA-generated alternatives and can process evaluation data from the optimisation model in different units of measurement. 
VF-MGA couples MGA and MCDA bidirectionally in three phases as illustrated in Fig.~\ref{fig:method_overview}. 

\subsection{Phase 1: MCDA problem structuring \& objective definition}
\label{subsec: VF-MGA_Phase1}
\textit{Phase 1} of VF-MGA builds on the established procedural structure of MCDA for problem structuring \citep{Belton.2002}. 
This includes the definition of the decision context \textit{(Step 1.1)} and the identification of relevant stakeholders \textit{(Step 1.2)}.
Relevant stakeholders may include everyone involved in the decision-making process, as well as anyone who may be affected by the outcome and therefore wish to have their interests considered.

VF-MGA is based on the principles of \textit{value-focused thinking} (VFT) \citep{Keeney.1992}. 
VFT shifts the focus from evaluating a given and potentially too narrow set of alternatives to identifying stakeholders' fundamental objectives.
Within VF-MGA, VFT supports the systematic identification of stakeholder-relevant alternatives in the subsequent MGA phase. \textit{Step 1.3} therefore focuses on the elicitation of stakeholders' objectives by means of structured interviews with previously identified stakeholders.
The objectives are organised in a hierarchical structure, ranging from broader high-level objectives to more specific low-level objectives \citep{Keeney.1992}. 
The objectives on the lowest level of the hierarchy are operationalised by defining attributes for which the alternatives' performance can be measured. 
Within VF-MGA, it is highly beneficial, to determine the performance for these attributes on the level of alternatives’ components, e.g., technologies or regions. To approximate these component-level contributions, additional information is required, including both the range of possible contributions and how attribute values scale with them. This may involve literature research, expert elicitation, or other data sources. Where possible, these components should be defined in a way that allows their alignment with the decision variables of the MGA optimisation model. In \textit{Phase 2}, the contribution of individual decision variables to attribute values is essential for guiding the MGA algorithm to generate stakeholder-relevant alternatives. In \textit{Phase 3}, determining attribute values from the large set of generated alternatives by aggregating component-level contributions, avoids repeated evaluation of individual alternatives (e.g. using separate models), thereby substantially reducing computational effort. 

Let \(P = \{p_1, \dots, p_n\}\) denote the set of attributes, while each objective is measured by one or more attributes. The performance of an alternative \(a_k\ \in A\) with respect to attribute \(p_j \in P\) is denoted by \(y_{kj}\).
The set of stakeholder objectives with their corresponding attributes derived in \textit{Phase 1} form the inputs to \textit{Phase 2}.

\subsection{Phase 2: Generation of decision alternatives with MGA}
\label{subsec: VF-MGA_Phase2}
\paragraph{Step 2.1: Model formulation \& MGA implementation}
\textit{Phase 2} starts with the development, implementation, and validation of the optimisation model. This includes the implementation of MGA, which we describe in more detail.
Starting from the original optimisation problem
\begin{equation}
\label{eq: MGA technique general}
\min_{x \in X} f(x),
\end{equation}
where $X$ denotes the feasible set defined by model constraints, MGA reformulates the objective function as a constraint and introduces the weighted sum of selected decision variables $x_i$, $i \in I$ as a new objective function: 
\begin{equation}
\label{eq:MGA weighted sum}
\min_{x \in X} \sum_{i \in I} w_i x_i 
\quad \text{s.t.} \quad
f(x) \le (1 + \varepsilon) f^* .
\end{equation}
Here, the slack parameter $\varepsilon$ defines the near-optimal solution space by constraining the relative deviation from the optimal objective value $f^*$ of the original optimisation problem (Eq.~\ref{eq: MGA technique general}).
Different choices of slack values $\varepsilon$ and weights \(w_i\) influence the size of the near-optimal solution space and the direction of structural diversification, respectively. By solving the optimisation problem (Eq.~\ref{eq:MGA weighted sum}) for different combinations of weights, slacks, and decision variables, a set of structurally distinct alternatives is generated. 

The choice of decision variables \(x_i\) determines the dimension along which alternatives are diversified (diversification dimension) and varies with the decision problem. For instance, \citet{Brill.1982} specify acreage allocated to land use \(i\) in region \(j\) as decision variables \(x_{ij}\). In energy system modelling, \citet{DeCarolis.2011} use energy generation or installed capacity as variables to generate structurally different energy system configurations.
More recently, \citet{Lombardi.2020} use spatially explicit variables to diversify the regional distribution of renewable energy technologies.
Moreover, groups of variables are often jointly considered in the MGA objective function instead of individual variables.
Let $G = \{G_1, G_2, \ldots\}$ denote a set of MGA groups, where each group \( G_m \subseteq I \) represents a subset of decision variables \(x_i\) that are jointly weighted in the MGA objective function.
Then all variables in the same group are assigned the same group-specific weight $w_m$, i.e.\ $w_i = w_m \ \forall \ i \in G_m$ in Eq.~(\ref{eq:MGA weighted sum}).

\paragraph{Step 2.2 Preference-informed MGA diversification}
Groups typically aggregate decision variables with obvious similarities, such as the same technology or region.
The rationale of such standard MGA approaches is that diversification along these groups and dimensions leads to the desired diversification.
It is not guaranteed, however, that these simple, modeller-driven decisions reflect stakeholder-relevant objectives.
The VF-MGA methodology proposed here systematically reflects stakeholder objectives derived from interviews in \emph{Phase 1} into the choice of MGA groups and their underlying diversification dimensions. The methodology is designed to generate alternatives that cover a wide range of attribute levels, thereby providing a meaningful basis for the subsequent preference elicitation in \emph{Phase 3}. It further aims to ensure a high diversity of outcomes within this range, reflecting varying objective trade-offs and increasing the likelihood that heterogeneous stakeholder interests are represented.
The sub-steps \textit{2.2.1 - 2.2.3} (cf.~Fig.~\ref{fig:method_overview}) are described in the following in more detail.

\paragraph{Step 2.2.1: Attribute-based MGA group construction} 
To reflect stakeholder objectives in the MGA diversification, each decision variable $x_i$, $i \in I$ is analysed with respect to its contribution to the attributes $p_j \in P$ that measure the objectives. The attributes are approximated through groups of decision variables with similar contributions to the respective attribute values (attribute-based MGA groups). For this purpose, component-level contributions to attribute values identified in \textit{Phase 1}, corresponding to decision variables in the MGA model, are required to assess the influence of decision variables on attribute values. Here, decision variables are distinguished based on whether they strongly drive an attribute value (driver groups \(G_{p_j}^{+} \subseteq I \)) or contribute to avoiding it (avoider groups \(G_{p_j}^{-} \subseteq I \)). Consequently, the number of MGA groups equals twice the number of attributes ($|G| = 2|P|$). Accordingly, the set of MGA groups $G$ becomes:

\begin{equation}
G = \{\, G_{p_j}^{+},\, G_{p_j}^{-} \mid p_j \in P \,\}.
\end{equation}

The diversification dimension of each attribute-based MGA group is implicitly determined by the type of decision variables driving the attribute values (e.g.~energy generation, installed capacity, spatial allocation). Both groups (driver and avoider) generate structural diversity concerning a given attribute by steering the respective contributing decision variables. Only attributes whose calculation can be meaningfully decomposed into contributions from individual decision variables are suitable for this MGA group construction. Attributes that do not allow such a decomposition (e.g.~due to interdependencies of decision variables) are not reflected through MGA groups and evaluated solely within the MCDA framework in \emph{Phase 3}. 

To systematically assign decision variables to the driver and avoider groups of each attribute, we propose two alternative assignment strategies. 
The \textit{contribution-based assignment strategy} groups at least two decision variables with the strongest contribution (driver groups) or lowest contribution (avoider groups) to the respective attribute.
The \textit{domain-balanced assignment strategy} is only relevant for models that cover multiple system domains, e.g.~energy sectors, with domain-specific decision variables. Since feasible alternatives require simultaneous adjustments across these domains, diversification restricted to the strongest contributors may affect only a subset of the system. The \textit{domain-balanced assignment strategy} ensures representation of all domains within each MGA group by assigning at least one decision variable per domain, thereby enabling system-wide diversification. Such domains may correspond to energy sectors, regions, land-use categories, or other model-specific components. 
For both assignment strategies, a relevance threshold should be applied to ensure that the selected decision variables are capable of exerting a meaningful influence on the system configuration. This is essential to ensure that changes in the decision variables translate into observable differences in the corresponding attribute values. This threshold is system-specific and should be chosen to avoid MGA groups composed of decision variables with only marginal system relevance while not requiring the inclusion of decision variables in each group that might dominate the group. In cases where multiple decision variables exhibit identical or very similar contributions to an attribute, they should be assigned collectively to the corresponding MGA group to ensure consistency and avoid arbitrary selection.

Within such groups, the variable-specific shift costs, i.e.~the specific objective value effect of increasing or decreasing the variable away from the optimum, drives which variable is diversified with priority. 
If attribute impact and shift cost do not align, the VF-MGA grouping may not work properly, as variables with lower shift-costs are favoured over those more relevant to the attribute value, particularly at low slack levels. 

\paragraph{Step 2.2.2: Application of weighting schemes} 
To explore the range of each attribute, at least one weighting schemes should be used that steers the optimisation towards extreme configurations.
A commonly used approach is extreme weighting, which successively maximises and minimises (groups of) decision variables within the specified slack $\varepsilon$ \citep{Neumann.2021}. The corresponding extreme weighting vector is defined as
\begin{equation}
\label{eq:extreme weighting}
w^{\text{E},\pm r}_i =
\begin{cases}
\pm 1, & \text{if } i = r, \\
0, & \text{otherwise},
\end{cases}
\quad \forall r \in I.
\end{equation}
In addition, VF-MGA can be combined with attribute-internal multi-extreme weighting adopted from the multi-extreme weighting by \citet{Esser.2024}.
Here, we assign opposite signs to the driver and avoider groups associated with the same attribute, while all other weights are zero
\begin{equation}
\label{eq:multi-extreme weighting}
w^{\mathrm{ME},\pm (r,s)}_i =
\begin{cases}
\pm1, & \text{if } i = r, \\
\mp1, & \text{if } i = s, \\
0, & \text{otherwise},
\end{cases}
\quad \forall r,s \in I, \; r \neq s.
\end{equation}
Additional weighting schemes like Hop-Skip-Jump \citep{Brill.1982} can be incorporated depending on the context.

\paragraph{Step 2.2.3: Application of slack levels \& alternative generation} After defining appropriate slack levels, decision alternatives are generated by solving the MGA optimisation problem (Eq.~\ref{eq: MGA technique general}) for different combinations of groups, weighting schemes, and slack levels.
This set of alternatives forms the decision basis for the subsequent MCDA evaluation.

\subsection{Phase 3: Preference elicitation and alternative evaluation}
\label{subsec: VF-MGA_Phase3}
In the third phase, the comprehensive set of alternatives obtained via MGA is evaluated according to the alternatives' performances on objectives and stakeholder preferences.
\paragraph{Step 3.1: Alternative performance prediction}
The performance of each alternative on each attribute is calculated or predicted. Depending on the optimisation model, some attribute values may already be computed during alternative generation (e.g.~cost indicators), such that \textit{Phase 2} can yield both decision alternatives and partial alternative performance information.
Further attributes defined at the component level can be calculated based on the collected data and specified aggregation rules in \textit{Phase 1}, and the composition of alternatives generated in \textit{Phase 2}. For other attributes, additional modelling may be required, particularly when their values depend on dynamic interactions between components of the alternatives. 
The evaluation process may involve iterative refinement, including revisiting \textit{Phase 1} to adjust attribute definitions or aggregation approaches, as well as \textit{Phase 2} to refine the assignment of decision variables if they do not adequately diversify attribute values.

\paragraph{Step 3.2: Stakeholder preference elicitation} In a second round of systematic stakeholder interaction, preferences from stakeholders are elicited. 
In general, preferences should be elicited from those stakeholders identified as relevant under \textit{Step 1.2} to ensure that the evaluation adequately incorporates the interests of a broad spectrum of stakeholders.
Incorporating elicited preference information over simplifying assumptions can have substantial effects on the results of the MCDA \citep[e.g.,][]{Scholten.2015, Haag.2019b}.
The chosen MCDA method determines the type and format of preference information that needs to be elicited and the structure of the model that is used to aggregate preferences with the alternatives' performances.  

\paragraph{Step 3.3: Decision model \& evaluation} A wide range of MCDA methods is generally suitable to be applied within VF-MGA. 
See \citet{Greco.2025} for a recent and general overview on common methods. 
However, we particularly emphasise multi-attribute value and utility theory (MAVT/MAUT) methods as they satisfy the requirements outlined above. 
MAVT is axiomatically grounded on preference axioms and provides a coherent method to operationalise the principles of VFT \citep{KeeneyRaiffa.1976, Keeney.1992}.
MAVT is particularly suitable to decisions in the environmental and public domain as it provides a transparent and structured framework for integrating heterogeneous objectives and making trade-offs explicit \citep{Reichert.2015, Greco.2016}.

As MAVT is adopted in this work, the performance of an alternative is evaluated at the attribute level using single-attribute value functions (SAVFs), denoted \(v_{j}(y_{kj})\).
SAVFs map the performance of an alternative to a dimensionless value between 0 and 1.
The shape of a SAVF represents the stakeholders' preferences for performance improvements at different attribute states. E.g., performance improvements might be valued higher (or lower) when the performance is comparatively bad.

Attribute weights, \(\mathrm{w}_j\),  with \(\mathrm{w}_j \geq 0\) and $\sum_{j=1} 
\mathrm{w}_j = 1$, model stakeholders' preferences for trade-offs between objectives.
Weights and SAVFs are combined within an aggregating function \(g\), which determines an overall performance score

\begin{equation}
V(a_k) = g(v_{1}(y_{k1}),v_2(y_{k2}), \dots, v_n(y_{kn}); \mathrm{w}_1, \mathrm{w}_2, \dots, \mathrm{w}_n)\
\end{equation}
for each alternative.
The resulting overall performance scores enable the comparison and ranking of alternatives, which informs decision-making.
The shape of the aggregation function models interdependencies between objectives, e.g., the allowed degree of compensation, and also depends on the preferences of the decision-maker. 
As the choice of an aggregation model can have considerable influence on the results of the MCDA it should be verified with the decision-makers \citep{Haag.2019}.

\section{Application of VF-MGA: A case study on the energy transition of a large university campus}
\label{chapter: case study}
In this section, we apply the VF-MGA methodology to a case study on the future decarbonised energy supply of a large university campus. We conduct an MCDA involving 11 stakeholder representatives and generate decarbonised energy supply alternatives with an ESM-MGA. The structure of this section follows the VF-MGA workflow illustrated in Fig.~\ref{fig:method_overview} and describes each step of the methodology in the context of the case study. 

\subsection{Phase 1: MCDA problem structuring \& objective definition}
\label{sec: MCDA before MGA}

\subsubsection{Decision-making context} 
We analyse the case of a large university campus, using the Ruhr University Bochum (RUB) in North Rhine-Westphalia (NRW), Germany, as an example. The decision-making context is framed by the climate protection act of the state of NRW that aims to remove anthropogenic greenhouse gas emissions in NRW by the year 2045 \citep{SVGNRW.01.04.2025}. The law states that universities should play a pioneering role in this effort. Beyond that, RUB itself aims to reduce its emissions by 2030 to 65\% and 100\% by 2045 compared to 1990.
Being one of the largest public research universities in Germany, with 37,600 students and 6,500 employees on an area of 4.5~km\textsuperscript{2}, RUB has a substantial energy demand \citep{AgenturderRUB.2023a}. In 2019, the total annual consumption amounted to 209 GWh, including 110 GWh for heating, 81 GWh for electricity, and 18 GWh for cooling. Currently, electricity for power consumption and cooling is mainly supplied by electricity procurement (EP). Heat supply is mainly covered by a company operating natural gas-based combined heat and power (CHP) plants. Future demand projections and renewable supply technologies to achieve the decarbonisation have already been studied by \citet{Esser.2024}. In this work, we investigate potential supply configurations for the year 2045. The corresponding demand assumptions and technology options considered are described in Section~\ref{subsec: Description ESM RUB}. The available technologies allows for a wide range of feasible supply configurations that we assess based on stakeholder preferences to identify preferred alternatives. In addition to the variety of available supply alternatives, a diverse range of stakeholders and institutions are involved in or affected by the decision-making process. 
These stakeholders have heterogeneous interests and preferences that need to be integrated into the decision-making process, highlighting the need to identify a consensus across differing perspectives. Against this background, the overall aim of this case study is the stakeholder-driven development and multi-criteria evaluation of feasible future energy supply concepts for RUB, while explicitly incorporating stakeholder preferences. 

\subsubsection{Identification of relevant stakeholders}
\label{subsubsec:stakeholder_identification}
The identification of relevant stakeholders follows a stratification and snowball sampling approach \citep[e.g., as in][]{Lienert.Schnetzer.Ingold.2013}. Initially, we compile a list of potentially relevant persons and groups. Subsequently, we verify and expand this list during the first round of semi-structured stakeholder interviews, using brainstorming enriched by the initial stratified sample. The participating stakeholders are summarised in Table~\ref{tab:stakeholders}.
In total, 16 stakeholders participate in the first round of individual interviews to identify relevant decision objectives (see Section~\ref{subsec: Identification of objectives}). Of these, 12 stakeholders also take part in the subsequent preference elicitation during the second round of individual interviews (see Section~\ref{sec: MCDA after MGA}). As two stakeholders jointly provide preferences due to shared decision-making responsibilities, we have 11 effective stakeholders in the second interview round. The reduced participation in the second round is primarily due to time constraints during the interview period.

\definecolor{headergray}{gray}{0.92}

\begin{table}[!t]
\centering
\scriptsize
\caption{Overview of stakeholder groups participating in the case study and numbers of participants per group. Objectives are elicited in the first and preferences in the second interview round. In total, 16 different individuals are involved.}
\label{tab:stakeholders}
\begin{tabularx}{\textwidth}{>{\raggedright\arraybackslash}X c c}
\toprule
& \multicolumn{2}{c}{Number of participants} \\
\cmidrule(lr){2-3}
Stakeholder group & Objective elicitation & Preference elicitation \\
\midrule

\rowcolor{headergray}
\multicolumn{3}{l}{\textbf{University administration / institutional management}}\\
\addlinespace[2pt]

Strategic university management
  & 2
  & 1 \\
    \addlinespace[2pt]

Energy management
  & 2
  & 1 \\
  \addlinespace[2pt]

Technical building management
  & 1
  & 1 \\
  \addlinespace[2pt]

Building operations
  & 1
  & 1 \\

\addlinespace[6pt]

\rowcolor{headergray}
\multicolumn{3}{l}{\textbf{University community}}\\
\addlinespace[2pt]

Student representation
  & 2
  & 2 \\
  \addlinespace[2pt]

Staff representation (academic, technical \& administrative)
  & 2
  & 2 \\

\addlinespace[6pt]

\rowcolor{headergray}
\multicolumn{3}{l}{\textbf{Owner of the building infrastructure}}\\
\addlinespace[2pt]

State-owned real estate agency (NRW)
  & 1
  & 1 \\

\addlinespace[6pt]

\rowcolor{headergray}
\multicolumn{3}{l}{\textbf{External service providers}}\\
\addlinespace[2pt]

Utility providers
  & 2
  & 0 \\

\addlinespace[6pt]

\rowcolor{headergray}
\multicolumn{3}{l}{\textbf{Academic experts}}\\
\addlinespace[2pt]

Energy experts
  & 2
  & 2 \\

\addlinespace[6pt]

\rowcolor{headergray}
\multicolumn{3}{l}{\textbf{Initiatives}}\\
\addlinespace[2pt]

Civil society initiative (climate-focused)
  & 1
  & 1 \\

\bottomrule
\end{tabularx}
\end{table}

\subsubsection{Identification of objectives}
\label{subsec: Identification of objectives}
We identify the decision relevant objectives in individual stakeholder interviews, following \citet{SchaerPreprint.2026}. In total, we interview 16 different stakeholders during this first round. The interviews focused on the underlying values and objectives of the stakeholders using brainstorming enriched by a master list of objectives to provide stimulative cues to stakeholders \citep{Haag.2019}.
We base the master list on three main sources of literature: studies examining similar decision contexts or problems, such as the sustainable energy supply for districts (e.g. \citet{Hottenroth.2022, Lerche.2019, HussainMirjat.2018}); studies already applying MCDA in university settings (e.g. \citet{Rajavelu.2021, Kriechbaum.2023, Lei.2025}); systematic review studies that specifically analyse and evaluate the choice of objectives in MCDA applications for renewable energy systems (e.g.~\citet{Wang.2009, Liu.2014}). Additionally, we incorporate non-scientific publications from RUB itself \citep{RuhrUniversitatBochum.2023}. 

During the first round of interviews, we continuously adapt and refine this list based on the stakeholders' input. Once all interviews are conducted and all objectives identified, we operationalise the rather abstract objectives to enable a systematic evaluation of RUB's future energy supply alternatives along the respective objective dimensions. To this end, we first structure the objectives into a hierarchy and then define attributes as performance measures to assess the achievement of objectives by alternatives \citep{Eisenfuhr.2010}. We specify the attributes in accordance with stakeholder feedback to ensure that they appropriately capture the meaning stakeholders assign to each objective.
The final objectives hierarchy is shown in Table~\ref{tab:objectives_hierarchy}. At the lowest level of the hierarchy, 11 attributes measure the performance of the energy supply alternatives with respect to the objectives.
In the supplementary information (SI-1) the objectives and their operationalisation are described in more detail. 

\definecolor{headergray}{gray}{0.92}  

\begin{table}[!t]
\centering
\scriptsize
\caption{Final objectives hierarchy constructed from individual stakeholder interviews. High-level objectives are printed in bold. Attributes measure the performance of each low-level objective. The direction specifies the preferred direction of the attribute value (down: lower is preferred; up: higher is preferred).}
\label{tab:objectives_hierarchy}
\begin{tabularx}{\textwidth}{>{\raggedright\arraybackslash}p{5cm}
                              >{\raggedright\arraybackslash}p{7cm}
                              l c}
\toprule
High- and low-level objective & Attribute & Unit & Direction \\
\midrule

\rowcolor{headergray}
\multicolumn{4}{l}{\textbf{1. Economic performance}}\\
\addlinespace[2pt]

{1.1 Operation and maintenance (O\&M) costs}
  & Annual O\&M costs
  & MEUR/a
  & $\downarrow$ \\
  \addlinespace[2pt]

{1.2 Investment costs}
  & Annual investment costs
  & MEUR/a
  & $\downarrow$ \\
  \addlinespace[2pt]

{1.3 Employee requirement}
  & Full-time equivalents (FTE) for operating energy supply technologies
  & FTE
  & $\downarrow$ \\

\addlinespace[6pt]

\rowcolor{headergray}
\multicolumn{4}{l}{\textbf{2. Environmental sustainability}}\\
\addlinespace[2pt]

{2.1 Primary energy factor (PEF)}
  & PEF of total energy supply
  & --
  & $\downarrow$ \\
  \addlinespace[2pt]

{2.2 Land-use-related environmental impact}
  & Land-use factor (LCA-based)
  & --
  & $\downarrow$ \\

\addlinespace[6pt]

\rowcolor{headergray}
\multicolumn{4}{l}{\textbf{3. Security of energy supply}}\\
\addlinespace[2pt]

{3.1 Exposure to energy price fluctuations}
  & Weighted price volatility exposure index
  & \%
  & $\downarrow$ \\
  \addlinespace[2pt]

{3.2 Energy diversity}
  & Shannon Index
  & --
  & $\uparrow$ \\

\addlinespace[6pt]

\rowcolor{headergray}
\multicolumn{4}{l}{\textbf{4. Feasibility of implementation}}\\
\addlinespace[2pt]

{4.1 Regulatory burdens}
  & Regulatory burden score
  & expert scale (1--7)
  & $\downarrow$ \\
  \addlinespace[2pt]

{4.2 Technical burdens}
  & Technical burden score
  & expert scale (1--7)
  & $\downarrow$ \\

\addlinespace[6pt]

\rowcolor{headergray}
\multicolumn{4}{l}{\textbf{5. Quality of stay on campus}}\\
\addlinespace[2pt]

{5.1 Campus area requirement}
  & Campus area requirement score
  & expert scale (1--7)
  & $\downarrow$ \\
  \addlinespace[2pt]

{5.2 Impact on campus operations}
  & Resource transport frequency
  & truck transports/a
  & $\downarrow$ \\

\bottomrule
\end{tabularx}
\end{table}

\subsection{Phase 2: Generation of decision alternatives with MGA}
\label{sec: ESM & MGA RUB}

\subsubsection{Energy system model of RUB: Model formulation and MGA implementation}
\label{subsec: Description ESM RUB}
We use an existing ESM of RUB \citep{Esser.2024}, implemented in the open-source framework \textit{Backbone} \citep{Helisto.2019}, to generate technically feasible, decarbonised energy system configurations for the subsequent MCDA. 
The cost function
\begin{equation}
\label{eq: optimization function backbone}
f_\text{cost} = \sum_{t} \omega_t \cdot ( c_t^\text{VOM} + c_t^\text{fuel} + c_t^\text{aux}) + c^\text{FOM} + c^\text{invest}
\end{equation}
measures total system costs, including annualised investment costs \(c^\text{invest}\), variable and fixed operation and maintenance costs \(c_t^\text{VOM}\) and \(c^\text{FOM}\), fuel and \(CO_2\) emission costs \(c_t^\text{fuel}\), and auxiliary costs  \(c_t^\text{aux}\). Costs at each time step \(t\) can be weighted with \(\omega_t\) to use representative periods.
The model is constrained by energy balance, energy generation and capacity limits, and emission restrictions. 
Technology investments and operational decisions are endogenously optimised by the model under exogenously specified energy demands, price assumptions, and local capacity constraints. 

The implementation of MGA constrains the cost function and minimises the weighted sum of grouped generation or capacity variables of energy generation or conversion units: 
\begin{equation}
\label{eq:MGA_Backbone}
\min  \Big(\sum_{G_m \in G} \Big(
  w^\text{invest}_{G_m} \sum_{i \in G_m} x_i^\mathrm{invest} +
  w^\text{gen}_{G_m} \sum_{i \in G_m} x_i^\mathrm{gen}
\Big) \Big)
\quad \text{s.t.} \quad
f_\text{cost} \leq (1 + \varepsilon) f^*_\text{cost} ,
\end{equation}
The MGA decision variables \(x_i\), \(i \in I\) represent the total energy generation or the investment capacity of energy system units, i.e., \(x_i^\mathrm{gen}\) or \(x_i^\mathrm{invest}\) in Eq.~(\ref{eq:MGA weighted sum}). They can be grouped into groups \(G_m \subseteq I\) by assigning the same weight to all units within each group, specified by the MGA weight parameters \(w^\text{invest}_{G_m}\) and \(w^\text{gen}_{G_m}\) for the investment and generation variables respectively. \(f^*_\text{cost}\) denotes the cost optimal objective value determined by Eq.~(\ref{eq: optimization function backbone}).
For further details, we refer to the original implementation by \citet{Esser.2024}.

The RUB energy system model covers heat, electricity, and cooling supply and uses time-series aggregation into representative weeks to reduce computational complexity. Electricity and cooling demands are projected to rise to 113 GWh and 30 GWh, respectively, while heating demand decreases to 103 GWh due to building modernisation \citep{Esser.2024}.
Heating technologies include biowaste-fuelled CHP (BioCHP), biomethane and pellet boilers, low- and high temperature heat pumps (LT-AWHPs, HT-AWHPs), a deep geothermal energy plant (DGE), and waste heat utilisation from RUB's data centre.
Furthermore, ground-water heat pumps (GWHPs) can be employed. Based on the available open area on the campus (approx. 0.6 km$^2$) and the required spacing between boreholes, a maximum of 1,030 geothermal probes is assumed \citep{Bracke.2015}, corresponding to a maximum geothermal heat potential of approximately 48 GWh per year. 
A total of 2.4 MW is already installed at the campus site and will be operating in the future. 
Cooling demand can be met by air-water heat pumps providing cooling (C-AWHPs), existing compression refrigeration machines (CRMs), and free cooling. Electricity is supplied by grid procurement (EP), with 0.875 MW of PV capacity planned for installation and potentially the BioCHP plant. The structure of the energy model is depicted in \ref{appendix: ESM specifications}.
Key techno-economic assumptions are summarised in the supplementary information (SI-3).

Moreover, we extended the MGA formulation to avoid the generation of weakly Pareto-optimal alternatives. 
Weak Pareto-optimality refers to solutions for which the constrained cost function value \(f_\text{cost}\) could be further reduced without increasing the weighted sum (cf.~Eq.~(\ref{eq:MGA weighted sum})). 
These solutions are dominated by solutions with the same diversity but lower cost, and thus inefficient and undesirable in the context of generating meaningful alternatives. 
To ensure the generation of Pareto-optimal solutions with MGA, the original formulation can be augmented analogously to the augmented $\varepsilon$-constraint method \citep{Mavrotas.2009} as introduced by \citet{Finke.2026}.

\subsubsection{Preference-informed MGA diversification}

\label{subsec: novel coupling approach MGA}
The preference-informed MGA diversification follows steps \textit{2.2.1--2.2.3} of the VF-MGA methodology (cf.~Fig.~\ref{fig:method_overview}).

\paragraph{Step 2.1.1: Attribute-based MGA group construction} 
For each attribute of the objective hierarchy elicited in the first round of stakeholder interviews (cf.~Table~\ref{tab:objectives_hierarchy}), we construct corresponding \textit{driver} and \textit{avoider groups} (\(G_{y_j}^+\), \(G_{y_j}^-\)) comprising technologies that exhibit a strong or low contribution to the respective attribute value.
As the attribute \textit{energy diversity} cannot be decomposed into technology-specific contributions, it is excluded from MGA group construction and solely evaluated within the MCDA. 
Moreover, regarding the attribute \textit{resource transport frequency}, only two technologies (pellet boiler and BioCHP) contribute to its value (cf.~Table~SI-2).
An \textit{avoider group} comprising all remaining technologies would not provide meaningful structural variation and is therefore omitted. Consequently, a total of 19 attribute-based MGA-groups are generated. 
In addition, we consider a conventional MGA grouping approach for comparison. In this grouping approach, we define a technology-benchmark MGA group set in which each technology forms its own group. Accordingly, each group contains exactly one technology, resulting in 13 technology-based MGA groups in total. 

The diversification dimension of each group is determined by the model variable driving the respective attribute value. In our ESM, this corresponds either to the capacity deployment or the energy generation. The calculation of each attribute, including whether it is defined on a capacity or generation basis, is summarised in Table SI-1.
With this, attribute-based MGA groups approximating investment costs, employee requirements, regulatory and technical burdens, and campus area requirements are defined as capacity-based MGA groups, while all others are defined as generation-based MGA groups. The technology-benchmark MGA groups are solely diversified along the energy generation dimension. 
The structured assignment of technologies to attribute-based MGA groups is based on their relative contribution to each attribute and implemented applying the two assignment strategies introduced in Section~\ref{subsec: VF-MGA_Phase2}, adopted to the specific characteristics of the case study.

Under the \textit{contribution-based assignment} strategy, the two highest-ranked technologies (driver groups) or lowest-ranked technologies (avoider groups) are selected first. To ensure that each group has a meaningful influence on the resulting alternatives, the assigned technologies must jointly be capable of covering a significant share of total energy demand. In this case study, a relevance threshold of 20\% is applied\footnote{\label{fn: explain_tech_threshold} This threshold is system-specific and depends on the considered system configuration. Several alternative values were tested to identify a suitable compromise: lower thresholds would allow groups with only marginal impact on objective outcomes, whereas higher thresholds would require the inclusion of technologies capable of supplying a substantial share of total demand, thereby potentially dominating the group effect.}.
If the initially selected technologies do not jointly meet this threshold, additional technologies are added in descending order of relevance until sufficient group-level impact is ensured.

The \textit{domain-balanced assignment} strategy is adapted to the three energy sectors considered in this study (electricity, heat, cooling). For each attribute-based MGA group, at least one technology per sector is selected whenever suitable options exist. Technologies are chosen by selecting the highest-ranked technologies (for driver groups) or lowest-ranked technologies (for avoider groups) within each sector.
To avoid negligible sectoral contributions, assigned technologies within a sector should jointly be capable of supplying approximately 40\% \textsuperscript{\ref{fn: explain_tech_threshold}} of the sectoral demand. In this work, this rule is applied exclusively to heat generation technologies. For electricity, procurement already covers at least 90\% of demand in all solutions and is therefore only included when crucial for influencing an objective. Due to the limited number of cooling technologies, this threshold is not meaningful for the cooling sector. 
For both assignment strategies, technologies with identical contributions are assigned collectively to the respective group, and identical technology sets are permitted to occur across different objectives.
The final MGA groups and the technologies assigned to them are listed in the supplementary information (SI-4).
In total, two assignment strategies are applied to 19 attribute-based MGA groups, resulting in 38 group variants, alongside 13 technology-benchmark MGA groups.

\paragraph{Step 2.2.2: Application of weighting schemes}
Group weights are applied following the extreme and attribute-internal multi-extreme weighting formulations introduced in Section~\ref{subsec: VF-MGA_Phase2} (cf.~Eq.~\ref{eq:extreme weighting}--\ref{eq:multi-extreme weighting}). For technology-based MGA groups, only extreme weighting is applied, resulting in 26 alternatives per slack value, as the attribute-internal multi-extreme weighting scheme does not provide additional structural insights for this group definition. For attribute-based MGA groups, both extreme and attribute-internal multi-extreme weighting schemes are implemented to explore a broader range of attribute-relevant trade-offs. Extreme weighting allows the individual performance of driver and avoider groups to be analysed separately, whereas attribute-internal multi-extreme weighting enables the simultaneous maximisation and minimisation of driver and avoider groups within the same attribute. This facilitates the investigation of interaction effects between both group types. As extreme weighting evaluates each MGA group twice (maximisation and minimisation), it effectively doubles the number of alternatives per MGA group and slack level. In contrast, multi-extreme weighting considers pairs of attribute-based MGA groups simultaneously, resulting in one alternative per attribute and slack level.

\paragraph{Step 2.2.3: Application of slack levels \& alternative generation}
According to \citet{Trutnevyte.2016}, up to 30\% cost slack is generally accepted. Therefore, we decided to use five different slack levels up to 30\% in this study (1\%, 5\%, 10\%, 20\%, and 30\%). 
We solve the MGA optimisation problem for all described combinations of MGA group definitions (attribute-based and technology-benchmark), weighting schemes (extreme and attribute-internal multi-extreme), and slack levels, thereby generating a total of $690$ alternative system configurations. Since the assignment strategies can lead to identical group assignments, identical alternatives arise across the strategies. Together with the cost-optimal alternative, this results in a total of $691$ alternatives that are subsequently evaluated within the MCDA. 

\subsection{Phase 3: Preference elicitation and alternative evaluation}
\label{sec: MCDA after MGA}
\subsubsection{Alternative performance prediction}
The performance of the 691 generated alternatives is evaluated with respect to each attribute, including uncertainty predictions.
Attribute values for O\&M costs and investment costs are directly derived from the ESM-MGA model. 
All other attribute values are calculated using the equations provided in Table~SI-1,
based on the underlying technology-specific contributions (see Table~SI-2)
and the composition of each alternative. As most attribute values can be derived through aggregation of technology-specific contributions with predefined coefficients, attribute values are calculated for the entire set of alternatives, without requiring additional models to evaluate individual alternatives.
To explicitly account for uncertainty in the performance prediction, we specify probability distributions for alternatives' performance on attributes. 
For all attributes except the three expert-based ones \textit{(regulatory burden}, \textit{technology burden}, \textit{campus area requirement}), we assume uncertainty to be normally distributed around the predicted mean attribute state (sd: $\pm 10\%$ of the mean). 
Uncertainty for predictions based on expert assessments is modelled via uniform distributions, based on the minimum and maximum expert estimates for each technology.
The best and worst predicted attribute states for each technology, including uncertainty, are used to calculate the impact range across all 691 alternatives. These ranges provide the basis for the subsequent preference elicitation. 

\subsubsection{Stakeholder preference elicitation and MCDA decision model building}
We elicit stakeholder preferences in a second round of individual interviews from each of the 11 participating stakeholders. 
During these interviews, we determine objective weights, assess the degree of compensation between objectives, and elicit the shape of single-attribute value functions (SAVFs) for the most important attributes within each high-level objective.
We adopt the elicitation protocols described in \citet{SchaerPreprint.2026}. 
An overview of the elicited parameters and used methods within this case study of RUB is provided in Table~\ref{tab:preference_elicitation}.

\begin{table}[h]
\centering
\scriptsize
\caption{Elicited stakeholder preference parameters and elicitation methods used to build the MCDA decision model.}
\label{tab:preference_elicitation}
\begin{tabularx}{\textwidth}{>{\raggedright\arraybackslash}X
                              >{\raggedright\arraybackslash}X}
\toprule
Preference parameter & Elicitation method\\
\midrule

Attribute weights (\(\mathrm{w}_j\))
  & SWING method \citep{vonWinterfeldt.1986, Eisenfuhr.2010}\\
  \addlinespace[2pt]

Single-attribute value functions (\(v_{j}(y_{kj})\))
  & Bisection method \citep{Eisenfuhr.2010}\\
  \addlinespace[2pt]

  Aggregation model (shape of aggregation function $\gamma$)
  & Indifference statements on hypothetical alternatives with different performances \citep{SchaerPreprint.2026}\\
  \addlinespace[2pt]
  
\bottomrule
\end{tabularx}
\end{table}

For a description of the widely applied SWING method to elicit attribute weights, we refer to the respective literature.

The elicited weights are then used to inform a time-efficient elicitation of SAVFs from each stakeholder. 
We approximate the shape of SAVFs for the attributes measuring the most important lowest-level objectives within each branch by eliciting mid-value points with the bisection method.
Based on the elicited mid-value points, the shape of the SAVFs is modelled via exponential interpolation.  
This provides a general idea on the shape of SAVFs for important objectives when stakeholders in practical applications are time-constrained.
However, we suggest to accompany such a time-conscious elicitation thorough sensitivity analysis.
We conduct all evaluations and sensitivity analyses using \textit{ValueDecisions} \citep{Haag.2022}, a web-based and open-source MAVT tool.

To determine the shape of the multi-attribute aggregation function, we use a simplified approach to test whether stakeholders agree to the preferential independence conditions that must be fulfilled to use a simple additive MCDA aggregation model.
We present them hypothetical alternatives with different performance levels across objectives and ask for indifference regarding the alternatives' overall value to test the acceptance of compensation between objectives.
Based on responses, we roughly approximate the parameter \(\gamma\) of a weighted power mean function as described by \citet{LANGHANS.2014}:

\begin{equation}
V(a_k) = M(v_{1}(y_{k1}),\ldots,v_{n}(y_{kn}); \mathrm{w}_1, \ldots, \mathrm{w}_n; \gamma) =
\begin{cases}
\left( \displaystyle \sum_{j=1}^{n} \mathrm{w}_j \, v_{j}(y_{kj})^{\gamma} \right)^{\frac{1}{\gamma}}
& \text{for } \gamma \neq 0 \\[10pt]
\displaystyle \prod_{j=1}^{n}v_{j}(y_{kj})^{\mathrm{w}_j}
& \text{for } \gamma = 0 
\end{cases}
\label{eq:power_mean}
\end{equation}
The weighted power mean aggregation function $M$ is particularly beneficial because it can represent vastly different compensation preferences through the one parameter $\gamma$.
For $\gamma = 1$, the weighted power mean function reduces to the additive model.
The overall value of an alternative, \(V(a_k)\), is used to rank alternatives for each stakeholder, derive stakeholder- and technology-specific insights and to identify consistently well-performing technologies.

\section{Results and discussion}
\label{chapter: results and discussion}
In this section, we present and discuss the results of the VF-MGA methodology applied to the decision context of RUB, including model results, stakeholder preferences, and the MCDA-based evaluation of alternatives.

\subsection{Range and diversity of modelling results}
\label{sec: ESM-MGA results}

\begin{figure}[p]
\centering
\includegraphics[width=1\textwidth]
{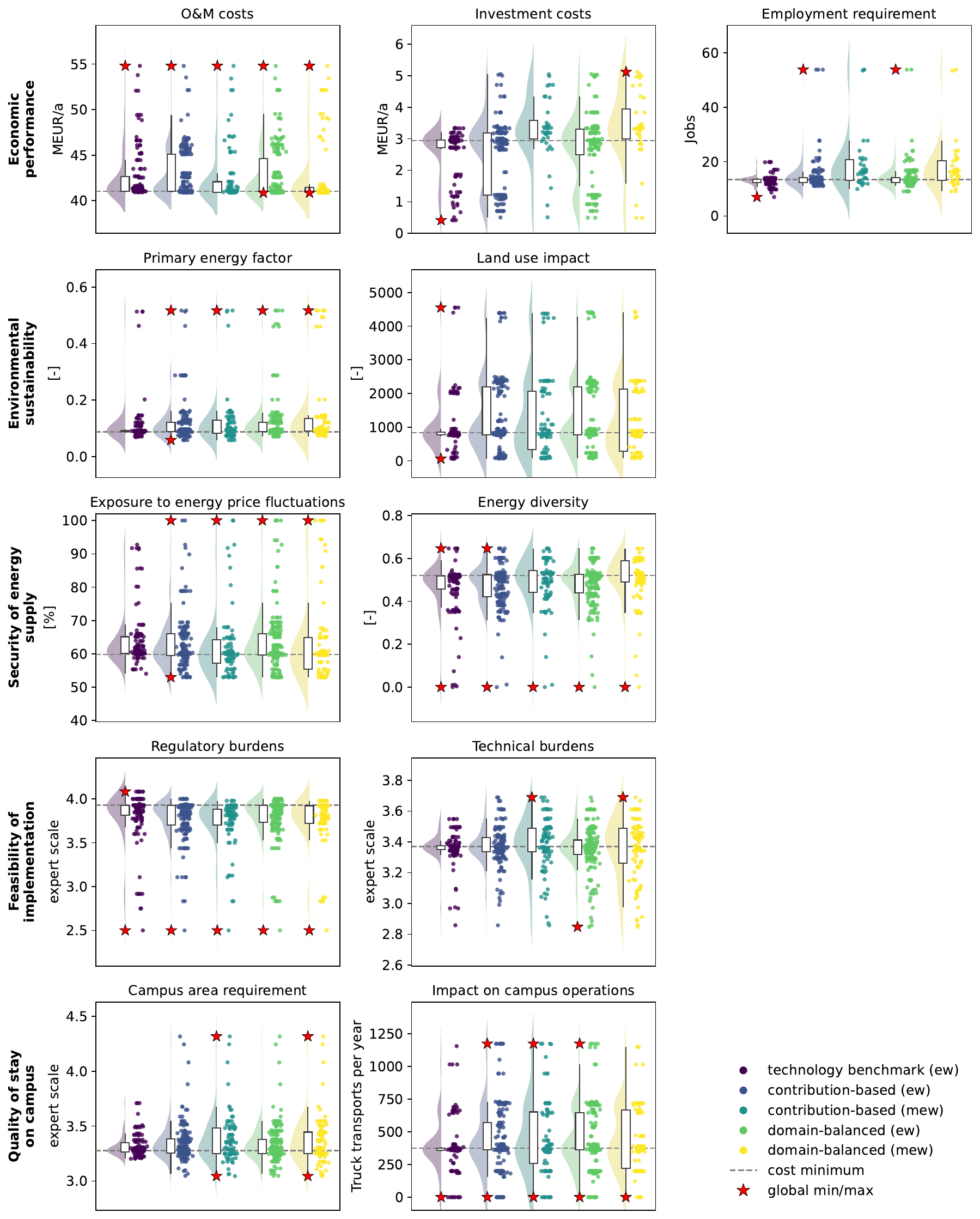}   
\caption{Attribute levels of the VF-MGA modelling results. Each facet displays the attribute levels (y-axis) for one objective, while colours present the grouping approach described in Section \ref{subsec: novel coupling approach MGA}. The labels ew (extreme weighting) and mew (attribute-internal multi-extreme weighting) denote the applied weighting schemes. Violin plots illustrate the density of the values (normalised), box plots show the dispersion, and scatter points indicate individual alternatives. Red stars mark the respective minimum and maximum values per objective. The dashed grey line indicates the cost-minimal solution.} 
\label{fig:MGA_rain_plot}
\end{figure} 

Fig.~\ref{fig:MGA_rain_plot} presents the range and diversity of the attribute values of all generated alternatives.
As compared to the technology-benchmark MGA groups, the VF-MGA grouping approach expands the range of attribute values significantly for several objectives (investment costs, employment requirements, technical burden, campus area requirement), all of which use capacity-based MGA groups\footnote{For the objectives \textit{investment costs} and \textit{employment requirements}, alternatives generated by maximising group variables with slack values above 5\% are excluded from Fig.~\ref{fig:MGA_rain_plot}, as they contain unused capacities that lead to very high objective values. This MGA artefact does not undermine the VF-MGA methodology, but could reduce the informative value of the figure, as discussed in more detail in the supplementary information (SI-5)}.
This indicates that VF-MGA may improve the performance of the MGA algorithm through choosing decision variables and their grouping in line with relevant decision objectives.
For the other attributes, the technology-benchmark spans (almost) across the entire range covered by the VF-MGA grouping approach.
This is enabled through a respective \emph{driver technology}, e.g.\ electricity procurement for O\&M cost or the pellet boiler for land-use impact, i.e.\ a technology with high specific impact on the attribute value and potentially high generation.
Therefore, grouping those drivers together with other technologies may even reduce the covered range, an effect that can be observed, albeit weakly, for land-use impact. 

While the applied weighting schemes are primarily designed to explore the boundaries of the near-optimal solution space, the VF-MGA grouping approach lead robustly to greater dispersion of attribute levels, reflected in wider interquartile ranges (IQRs) compared to the technology-benchmark groups. This effect can be attributed to the structure of the VF-MGA grouping approach. Each attribute is represented by a driver and an avoider group, generating alternatives that emphasise different levels of the same attribute. This promotes broader coverage of the attribute range beyond pure boundary exploration and results in a more diverse set of alternatives that better reflect heterogeneous stakeholder preferences. 
However, the increase in dispersion varies across attributes. This can be explained by the extent to which the technologies driving a given attribute can vary. For instance, land-use impact and impact on campus operation show a substantially larger IQR, as they are strongly influenced by technologies with high variability across alternatives (e.g.~pellet boiler). In contrast, for instance, the primary energy factor (PEF) is mainly driven by the biomethane boiler, which is generally used only for smaller heat generation volumes across alternatives, likely due to its high O\&M costs, resulting in comparatively lower dispersion of PEF values. A smaller increase in dispersion may also result from the mathematical formulation of the attribute, as aggregation or averaging effects can dampen variations in technology composition.

When examining the differences between the VF-MGA assignment strategies and their respective weighting schemes, no single strategy consistently outperforms the others. This is likely related to the model structure, in which most variation is driven by the heating sector, while the electricity and cooling sectors offer comparatively limited flexibility, resulting in similar diversification patterns across both assignment strategies. Attribute-internal multi-extreme weighting appears particularly promising, as it enables a more explicit steering along stakeholder-relevant attributes. 

\subsection{Stakeholder preferences}
\label{sec: Stakeholders' preference information}

\begin{figure}[!htb]
\centering
\includegraphics[width = 1\textwidth]{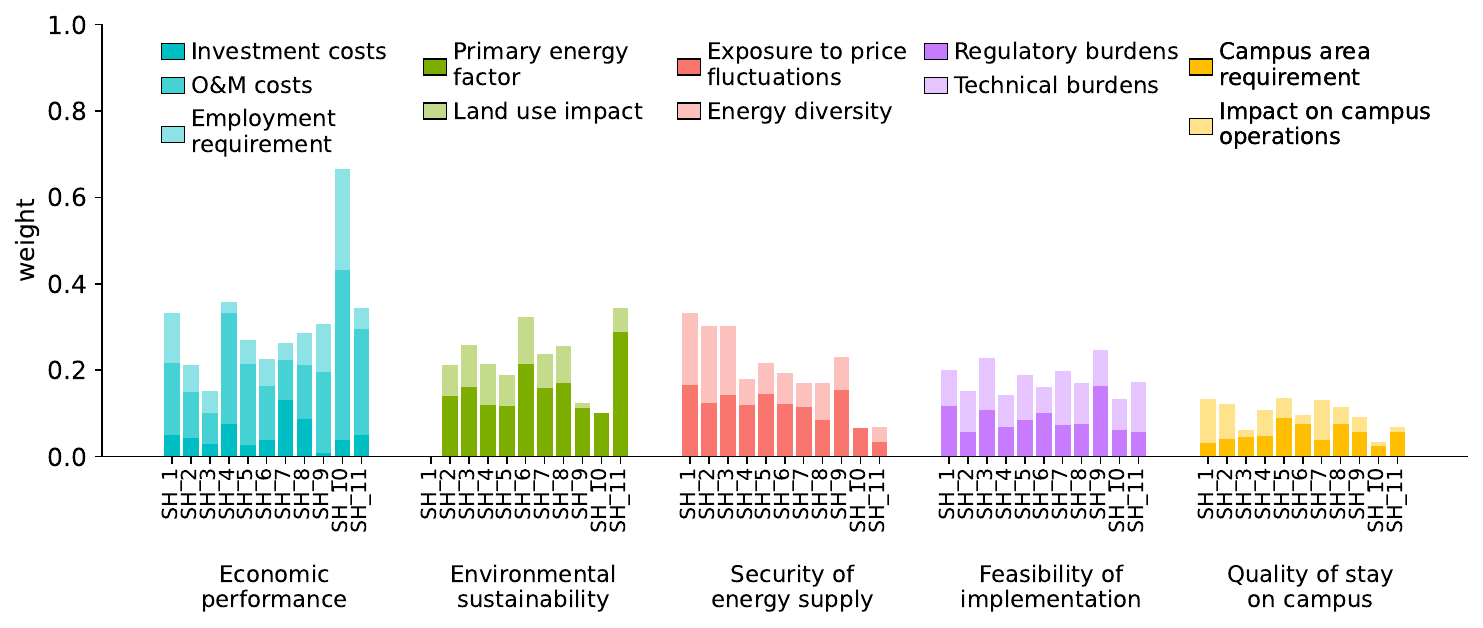} 
\caption{Weights (y-axis) for eleven low-level objectives grouped in high-level objectives (colours) as elicited in the interviews from eleven stakeholders. To ensure data protection, all stakeholders (SH) are referred to in anonymised form. Weights for each low-level objective are combined to a high-level objective weight (stacked bars). Bar segments with varying shades of colour indicate weights of low-level objectives. All weights per stakeholder sum up to one.} 
\label{fig:weight plot}
\end{figure}

Fig.~\ref{fig:weight plot} displays the weights for both low-level and high-level objectives for each stakeholder (SH) as elicited in the second round of interviews, revealing substantial heterogeneity in stakeholder preferences at both levels.
Beyond this heterogeneity, there is some consensus, e.g.~that economic performance is regarded as the most important and quality of stay on campus as the least important high-level objective by most stakeholders.
Nonetheless, economic performance, environmental sustainability, and security of energy supply exhibit the greatest variation in weights across stakeholders, making them the main drivers of preference heterogeneity. 
This is further illustrated by individual stakeholder perspectives. SH\_10 assigned a weight of 0\% to energy diversity, arguing that increased energy diversity may imply greater technological diversity, which is not considered desirable for the campus due to higher personnel requirements. Similarly, SH\_11
emphasised that energy diversity alone is insufficient and should be complemented by
supplier diversity, leading to a comparatively low weight. In addition, some stakeholders pointed out that fluctuating electricity prices are not necessarily disadvantageous, as long as there are appropriate technologies and strategies available to manage them.
In contrast, some preferences show relatively little variation across stakeholders. The feasibility of implementation exhibits comparatively small variation at the higher level and remains relatively balanced across different weightings of regulatory and technical burden.
Moreover, O\&M costs consistently represent the most important low-level objective within the economic performance (except for SH\_7), and within environmental sustainability, the PEF is consistently considered more important than land-use impact. 
SH\_1 chose not to assign weights to the environmental objectives, stating that due to extensive professional experience in the field, it is difficult to provide generalised preference statements based on the selected attributes. 
Overall, the key objective dimensions driving stakeholder differentiation in this analysis are economic performance, environmental sustainability, and security of supply, as they combine high relative importance with significant heterogeneity across stakeholders.

The shapes of the elicited single-attribute value functions (SAVFs) vary to a considerable extent across stakeholders, reflecting heterogeneous preferences. With the exception of the PEF and a small number of individual cases, most stakeholders place greater emphasis on performance improvements when performance is poor, leading to mostly concave SAVFs. The shape of the elicited SAVFs per stakeholder is shown in Table~SI-7. 
For the remaining low-level objectives, we assume linear value functions.

Finally, stakeholders largely rejected the implications imposed by an additive aggregation model. 
These preferences indicate a low degree of compensation, which we reflect by setting ($\gamma = 0.2$) as baseline, and assess the robustness of this modelling choice via sensitivity analyses in the supplementary information (SI-10).

\subsection{Ranking of alternatives}
\label{sec: Ranking of alternatives}

\begin{figure}[!htb]
\centering
\includegraphics[width = 1\textwidth]{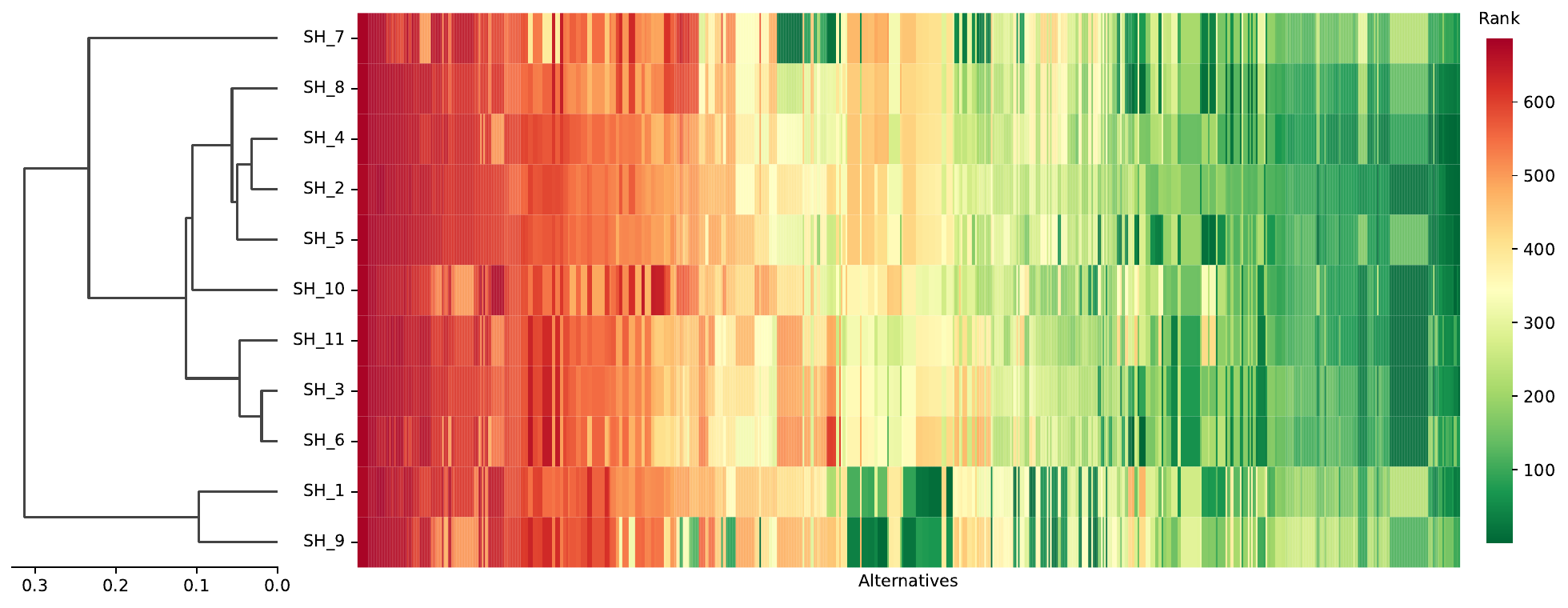} 
\caption{Heat map of stakeholder rankings for all alternatives, where colours indicate the rank of each alternative and rank = 1 denotes the best-performing one. Alternatives are ordered according to their mean rank across stakeholders. The dendrogram on the left shows hierarchical clustering of stakeholders based on Spearman rank correlation of their alternative rankings, such that smaller distances indicate higher similarity in preferences between stakeholders.} \label{fig:HeatsMap_Rank}
\end{figure}

The alternatives' performances across objectives are aggregated using stakeholder preferences to derive stakeholder-specific rankings of all 691 alternatives obtained with the ESM-MGA model.
The heat map in Fig.~\ref{fig:HeatsMap_Rank} illustrates the stakeholder-specific ranking patterns across all alternatives, revealing substantial variation in rankings across stakeholders. 
This indicates the absence of a best-performing alternative across all stakeholders and, consequently, the absence of a clear consensus.
Additionally, the dendrogram in Fig.~\ref{fig:HeatsMap_Rank} reveals that preferences lead to distinct ranking structures among stakeholders. In particular, SH\_1 and SH\_9 form a subgroup with similar rankings, while SH\_7 forms a distinct subgroup. Both exhibit substantial ranking dissimilarity from the remaining stakeholders.
SH\_1 and SH\_9 assign relatively low weights to environmental sustainability while simultaneously assigning higher weights to security of supply. SH\_7, in contrast, assigns comparatively high importance to investment costs, which may contribute to its distinct position. The ranking patterns of the remaining stakeholders appear more homogeneous despite the heterogeneous weight structures. Notably, these groups do not align with the predefined stakeholder groups (cf.~Table \ref{tab:stakeholders}), but instead have heterogeneous stakeholder compositions.

An examination of the total values for each alternative and stakeholder (cf.~Fig.~SI-3) reveals that several alternatives achieve comparably high values and thus perform similarly well for each stakeholder. This motivates analysing a broader set of best-ranked alternatives in the following rather than few individual solutions. 

The cost-optimal alternative ranks between rank 50 and 200 across stakeholders, underlining that the cost-optimal alternative that is typically obtained by ESMs without MGA does not necessarily align with stakeholder preferences.
Therefore, several near-optimal alternatives perform substantially better when evaluated against a broader set of objectives under the consideration of stakeholder preferences.
Stakeholders seem to accept additional costs for improving other objectives, which is in line with findings from related literature \citep{Vagero.2025, Trutnevyte.2016, TomasPascual.2025}.

\subsection{Value-focused must-haves and real-choices}
\label{sec: Value-focused must-haves and choices}
\begin{figure}
    \centering
    \includegraphics[width=0.55\linewidth]{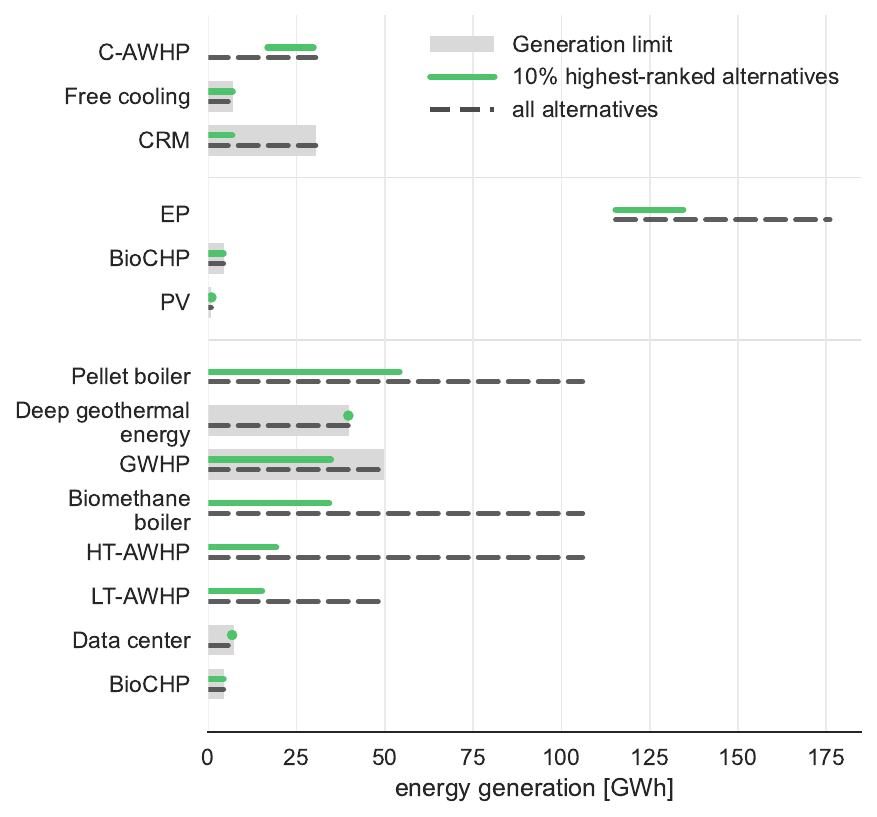}
    \caption{Energy generation variability across alternatives. Dark dashed lines represent the full set of generated alternatives, while green solid lines indicate the range of the 10\% highest-ranked alternatives for each stakeholder. Technical generation limits imposed in the ESM are illustrated by grey boxes.}
    \label{subfig: energy_gen_top10}
\end{figure}

Across all 691 alternatives generated with MGA, the possible deployment levels of most technologies span a wide range, as illustrated by the full generation ranges shown in Fig.~\ref{subfig: energy_gen_top10}. 
With the exception of the LT-AWHP, all technologies can be deployed across their entire feasible generation range. For capacity-constrained technologies, this range is limited by their maximum generation capacity, whereas for unconstrained technologies it is effectively determined by the respective total sectoral demand. Consequently, apart from EP, which constitutes a trivial \textit{must-have} technology in the absence of alternatives, no technologies can be identified as clear \textit{must-have} or \textit{must-avoid} technologies, but all appear as \textit{real-choice} options. \textit{Must-have} technologies denote those present in all feasible alternatives, \textit{must-avoid} technologies those absent from all alternatives, \textit{real-choice} technologies those appearing only in a subset of configurations with varying contribution.
Deriving a meaningful decision solely from this large set of near-optimal system configurations is therefore challenging.

Within VF-MGA, the preference-based ranking of alternatives from MCDA is used to reduce the solution space to alternatives that perform comparatively well for each stakeholder. By evaluating the alternatives with respect to stakeholder preferences, the range of relevant configurations can be narrowed considerably, as illustrated by the green dashed lines in Fig.~\ref{subfig: energy_gen_top10}. We focus on the 10\% best-ranked alternatives per stakeholder to reveal the most relevant part of the solution space.

Across all stakeholders, the deep geothermal energy plant (DGE) and data centre always operate at their maximum capacity within these top-ranked alternatives. This suggests that these technologies form a consistent component of stakeholder-preferred system configurations and constitute \textit{value-focused must-have} technologies, i.e., technologies that systematically appear in all preferred alternatives. 
The DGE and data centre combine comparatively low O\&M costs, which also contribute to reduced exposure to price fluctuations and higher energy diversity (see supplementary information (SI-6) 
for a correlation analysis of objectives), with strong performance across environmental sustainability objectives. This finding is particularly interesting in light of the trade-off between investment costs and O\&M costs (cf.~SI-6).
Despite its comparatively high investment costs, the DGE remains a \textit{value-focused must-have} technology for all stakeholders. This suggests that higher investment costs might be acceptable when technologies perform particularly well in other key objective dimensions, notably security of supply and environmental sustainability, where the PEF plays a key role.
For cooling technologies, the C-AWHP also emerges as a \textit{value-focused must-have} technology, although with a varying deployment range covering at least 56\% of the cooling demand, probably related to its higher efficiency and thus lower O\&M costs. EP is required to cover at least 90\% of the electricity demand due to limited alternative supply options. However, within this constraint, the focus on the 10\% best-ranked alternatives narrows the preferred range of EP by around 70\%, favouring configurations that avoid excessively high electricity demand. 

In contrast, the remaining heating technologies are not consistently present among the top-ranked alternatives. They seem to be \textit{value-focused real-choice} options, and their inclusion appears to depend on the stakeholder perspective. However, the feasible range of choices can be substantially reduced. For the CRM, biomethane boiler, HT-AWHP, and LT-AWHP the \textit{real-choice} range decreases by more than 65\%, reaching up to approximately 82\% for the HT-AWHP, thereby considerably narrowing the decision space towards stakeholder-preferred ranges. Notably, the technologies exhibiting the strongest reductions are also associated with comparatively high O\&M costs, which in turn, are correlated with other objectives (cf.~SI-6).
Moreover, no technology can be categorised as systematically unfavourable; hence, no \textit{value-focused must-avoid} technology exists in the investigated system. This likely reflects the heterogeneity of stakeholder preferences, as each technology performs well under at least some preference configurations.

Overall, MGA yields a broad set of near-optimal alternatives with no clear \textit{must-have} technologies, limiting its effectiveness for decision support. By incorporating stakeholder preferences, VF-MGA narrows the solution space substantially, revealing \textit{value-focused must-have} technologies and considerably reducing the range of \textit{value-focused real-choice} options. This leads to a more focused set of decision-relevant alternatives that better supports targeted decision-making.

\subsection{Consensus on technology deployment}
\label{sec: Consensus on deploying technologies}

Examining the 10\% best-ranked alternatives for each stakeholder reveals that no alternative is included within all stakeholder-specific subsets, emphasising a lack of clear consensus. The cost-optimal alternative is only present in the 10\% best-ranked alternatives for three stakeholders (2, 10, 11). 
We therefore analyse the \textit{value-focused real-choice} heating technologies in greater detail to gain a more differentiated understanding of stakeholder preference patterns and potential consensus. A corresponding analysis for cooling technologies and EP is provided in the supplementary information (SI-9).
Fig.~\ref{subfig:frequency_top10} illustrates the occurrence frequencies of heating technologies in the 10\% best-ranked alternatives, highlighting substantial variation across stakeholders.
In particular, the LT-AWHP, HT-AWHP, and the biomethane boiler are not consistently included in the best-ranked alternatives.
They are generally associated with higher O\&M costs, increased exposure to price fluctuations, or increased PEF values. As these objectives receive comparatively high weights from certain stakeholders, their inclusion can substantially reduce the aggregated value of an alternative from those perspectives.
Consequently, these technologies may represent potential points of conflict within the decision process, complicating consensus formation.
For the BioCHP, pellet boiler, and GWHP, a broader consensus can be observed. Around eight stakeholders consistently use these technologies in all of their 10\% best-ranked alternatives, whereas two to three also favour configurations without them. 

\begin{figure}
    \centering
    \includegraphics[width=0.6\linewidth]{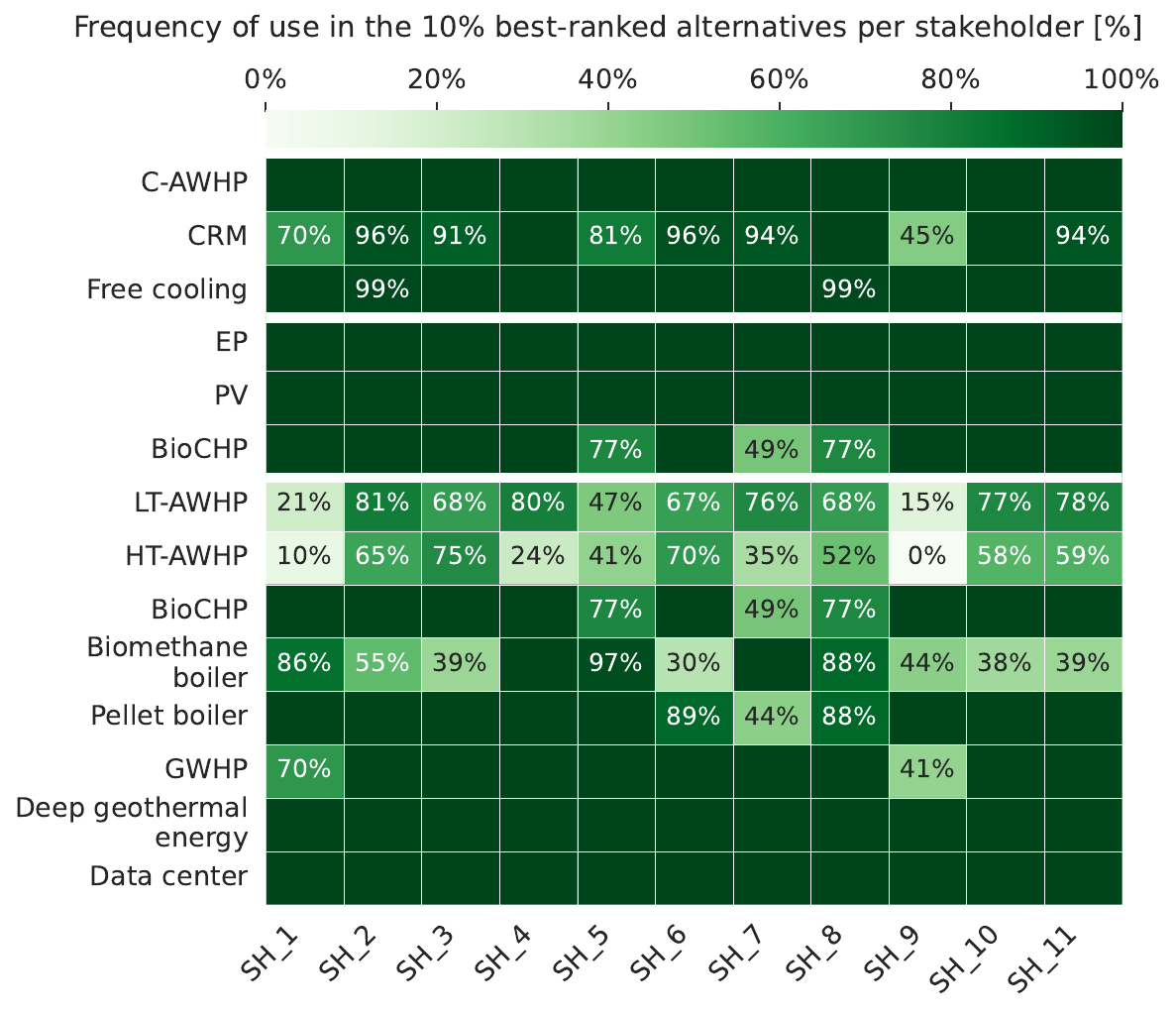}
    \caption{Heat map illustrating the normalised frequency of use per technology and stakeholder within their respective 10\% best-ranked alternatives. Only values below 100\% are labelled.}
    \label{subfig:frequency_top10}
\end{figure}

\begin{figure}[!htb]
    \centering
    \includegraphics[width=\linewidth]{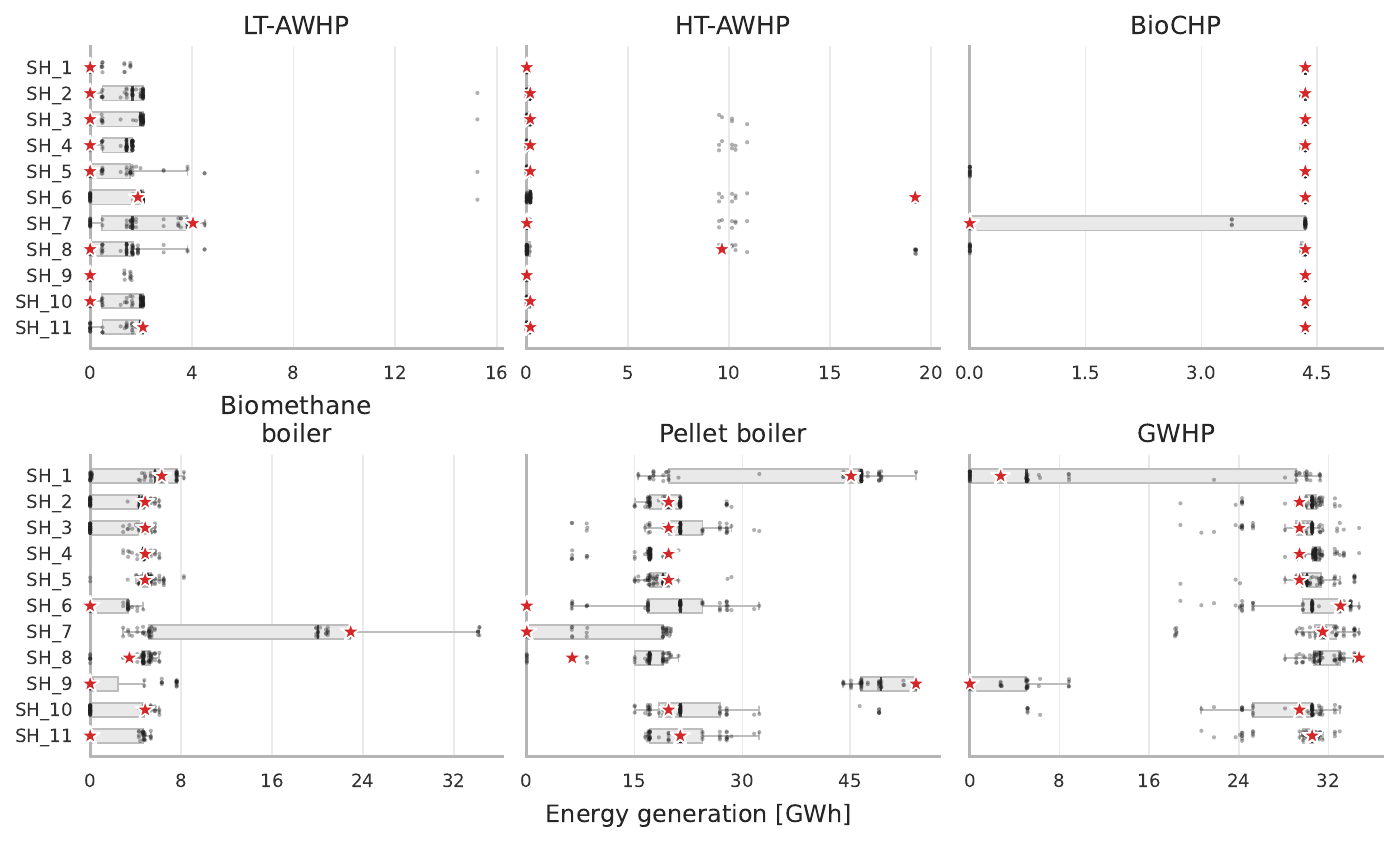}
    \caption{Energy generation from heating technologies across the 10\% best-ranked alternatives per stakeholder. Box plots and scatter distributions reveal the energy generation per technology across stakeholders' preferred alternatives. Red stars indicate the generation levels in the best performing alternative for each stakeholder.}
    \label{fig:boxplot_heat_top10}
\end{figure}

To further analyse consensus on technology deployment, Fig.~\ref{fig:boxplot_heat_top10} shows the heat generation of each \textit{value-focused real-choice} heating technology across the 10\% best-ranked alternatives per stakeholder.
Several disagreements between stakeholders become clear.
First, stakeholder 7's best-ranked alternatives include substantially higher shares of the biomethane boiler compared to those of the other stakeholders. This preference may be linked to the comparatively low investment costs of the biomethane boiler, an objective to which Stakeholder 7 assigns substantial weight. This may also explain the comparatively low utilisation of BioCHP in Stakeholder 7’s top-ranked alternatives, given its higher capital requirements.
In contrast, BioCHP is used in all configurations preferred by all other stakeholders, probably due to its contribution to energy diversity and its comparatively favourable performance regarding the PEF and exposure to price fluctuations.
Second, most stakeholders agree on installing substantial GWHP capacities on campus. Notable exceptions are stakeholders 1 and 9, who assign a relatively high importance to minimising exposure to price fluctuations and therefore favour configurations with lower electricity dependency. This is also reflected in the low occurrence of HT-AWHP  (0\% and 10\%, respectively) and LT-AWHP (21\% and 15\%, respectively; cf.~Fig.~\ref{subfig:frequency_top10}) within their 10\% best-ranked alternatives. At the same time, both stakeholders assign low importance to land-use impact, which diminishes the penalisation of biomass-based technologies that are present to some extent for many other stakeholders. The combined weighting structure (high weight on exposure to price fluctuations and low weight on land-use impact) thus increases the relative attractiveness of the pellet boiler compared to other stakeholder perspectives. 
Finally, stakeholders 3, 6, 7, and 8 assign comparatively low weights to O\&M costs and to security of supply objectives while simultaneously placing greater emphasis on environmental sustainability. This weighting pattern reduces the penalisation of electricity-dependent technologies and increases the relative importance of environmental performance, resulting in a higher acceptance of HT-AWHP. Accordingly, Stakeholders 6 and 8 accept substantial HT-AWHP capacities in their highest-ranked alternatives.

In summary, stakeholder disagreement primarily concerns the deployment levels of HT-AWHP, LT-AWHP, biomethane boiler, pellet boiler, and GWHP. This disagreement reflects both the trade-off between investment and O\&M costs, and differing preferences regarding whether the remaining heat demand should be met through electricity-based or biomass-based technologies. These differences can largely be attributed to the varying weight assignments in the key objective dimensions: economic performance, environmental sustainability, and security of supply. Objectives related to the feasibility of implementation and quality of stay on campus have comparatively little influence on the major preference conflicts observed across stakeholders and do not appear to be decisive drivers of disagreement. Given these differences, a structured joint workshop with all stakeholders could support the transparent discussion of underlying value trade-offs, increase mutual understanding, and facilitate the identification of compromise solutions. 

The patterns described above illustrate an important benefit of the VF-MGA methodology. Since multiple alternatives achieve similarly high value scores, examining this broader set of best-ranked alternatives allows for the identification of stakeholder-specific preferred ranges of technology deployment. By systematically generating structurally distinct alternatives, the VF-MGA-based exploration of the solution space provides more nuanced insights into stakeholder preferences than MCDA analyses based on a small set of alternatives. Rather than identifying a single preferred configuration, this reveals the range of solutions that stakeholders consider acceptable and helps identify potential areas of agreement between stakeholders. 
  
\subsection{Limitations \& outlook}
\label{sec: Limitations and outlook}

The proposed VF-MGA methodology and the case study have several limitations. 
Regarding the VF-MGA methodology, stakeholder-relevant objectives are approximated through groups of decision variables in the MGA model. However, as the resulting attribute levels of these groups depend on various model parameters and system interactions, the grouping approaches may not fully capture the underlying stakeholder objectives. Future work could therefore combine VF-MGA with multi-objective optimisation to capture selected objectives rigorously through objective functions instead of groups, where appropriate.

Additionally, the construction of attribute-based MGA groups relies on heuristic assignment strategies. As the effectiveness of these strategies may depend on the specific system configuration, they may require adjustment or further refinement when applied to other case studies. Furthermore, future research could develop decision variable-specific shift cost indicators that capture the alignment of attribute impacts of decision variables and their corresponding shift costs to refine the heuristic assignment strategies proposed in this work. 

Moreover, some objectives depend on the overall system configuration of an alternative and cannot be decomposed into technology-specific contributions, like energy diversity in our case study. As the proposed grouping approaches approximate stakeholder-relevant objectives with MGA groups, such objectives cannot directly guide the MGA diversification but can only be evaluated ex post within the MCDA.

Finally, the application of MCDA methods typically designed for small sets of alternatives, to a large set with hundreds of model-generated alternatives presents methodological challenges.
We therefore analysed patterns across sets of high-performing alternatives instead of focusing on individual alternatives. Nevertheless, systematic MCDA-based approaches for exploring and structuring preferred configurations within large MGA solution spaces could further improve such analyses.

Beyond these methodological limitations, further ones relate to our specific case study. 
First, technical system reliability was not explicitly modelled within the ESM-MGA and could not be considered within the MCDA evaluation, although several stakeholders identified it as an important objective (see SI-1 for further discussion). In addition, stakeholders mentioned further potential technologies and infrastructure options during the interviews, such as the use of waste heat from a nearby wastewater treatment plant or the installation of a thermal storage facility. Due to missing techno-economic data, these options could not be included in the present analysis. Once such information become available, a reassessment of the system configuration may provide additional insights.

\section{Conclusion}
\label{chapter: Conclusions and Outlook}
Identifying and evaluating decision alternatives in complex planning settings involving diverse stakeholders and multiple, often conflicting objectives remains a challenging task for which guidance is needed. To address this, we introduce \textit{value-focused modelling to generate alternatives (VF-MGA)}, a novel methodology that bidirectionally couples optimising models applying modelling to generate alternatives (MGA) and multi-criteria decision analysis (MCDA), enabling stakeholder-guided exploration and evaluation of near-optimal system configurations. As energy systems transformation exemplifies such complex decision contexts, we demonstrate \textit{VF-MGA} for the decarbonisation of a large university campus and involve 11 stakeholder representatives.
Several key advantages emerge from the application of the proposed methodology.

\textit{First}, stakeholder objectives elicited through MCDA can provide meaningful guidance for modelling decisions within MGA. This concerns both the grouping of decision variables and the system dimensions along which alternatives are diversified. The proposed VF-MGA approach thereby improves the representation of stakeholder-relevant objective values compared to the technology-benchmark groups through expanded value ranges for some attributes and consistently greater dispersion of attribute levels.
In particular, approximating stakeholder objectives through several structurally distinct groups, such as driver and avoider groups, appears beneficial, as it promotes alternatives emphasising different levels of the same attribute. Taken together, using objectives to guide the design and diversification of the MGA algorithm can improve the exploration of stakeholder-relevant regions in the solution space. 

\textit{Second}, evaluating the generated MGA alternatives using stakeholder objectives and preferences elicited through MCDA proves to be particularly valuable, as it enables the systematic comparison and prioritisation of alternatives, thereby narrowing the large MGA solution space to only decision-relevant configurations.

\textit{Third}, analysing a large set of structurally distinct alternatives within the MCDA reveals multiple system configurations that perform similarly well from the perspective of each stakeholder. This allows for the identification of acceptable ranges of decision variables or system components for individual stakeholders. Comparing these ranges across stakeholders reveals where acceptable levels overlap, indicating potential areas of agreement, and where they diverge, highlighting sources that give rise to genuine stakeholder conflicts. This enables a comprehensive understanding of stakeholder preferences and provides a transparent basis for identifying potential compromise solutions.

Finally, we demonstrate in a case study with real stakeholders that the cost-optimal alternative does not necessarily align with stakeholder preferences and that stakeholders seem to accept additional costs for improving other objectives. This highlights the importance of exploring near-optimal alternatives, as enabled by MGA, and explicitly incorporating stakeholder preferences in complex planning processes. 

To sum up, VF-MGA links the exploration of near-optimal system configurations with a structured elicitation and evaluation of stakeholder preferences. By guiding the diversification of alternatives through stakeholder objectives and subsequently evaluating a large set of model-generated alternatives within an MCDA framework, VF-MGA bridges the gap between optimisation-based modelling and decision-orientated analysis. In this way, the proposed methodology enables the exploration of feasible solution spaces complemented by an assessment of what is desirable from a stakeholder perspective. VF-MGA therefore provides a systematic methodology for translating modelling results into decision-relevant insights for complex planning decisions. 

Regarding the conducted case study at RUB, the analysis indicates that the most relevant objective dimensions are economic performance, ecological sustainability, and security of supply. The strongest disagreement between stakeholders arises for technologies that affect these dimensions, particularly bio-based technologies and technologies with considerable electricity demand, which are associated with high energy costs, strong price volatility, or considerable environmental impacts. In contrast, a high degree of agreement among stakeholders can be observed for technologies that have only limited impacts across these key objective dimensions. This includes the DGE and the data centre, whose deployment levels show comparatively little influence on the key stakeholder objectives. 

These findings illustrate how VF-MGA can help identify technologies that drive potential stakeholder conflicts, as well as those that are broadly acceptable across stakeholders, thereby supporting more informed and transparent energy system decision-making.

\paragraph{Author contributions}
\textbf{Emily Bergup:} Writing – review \& editing, Writing – original draft, Visualisation, Software, Methodology, Investigation, Formal analysis, Data curation, Conceptualisation.
\textbf{Jonas Finke:} Writing – review \& editing, Writing – original draft, Supervision, Software, Methodology, Conceptualisation.
\textbf{Sebastian Schär:} Writing – review \& editing, Writing – original draft, Supervision, Software, Methodology, Conceptualisation.
\textbf{Valentin Bertsch:} Writing – review \& editing, Supervision, Conceptualisation.

\paragraph{Acknowledgements}

We sincerely thank all stakeholders for their generous support and willingness to participate in this study. We greatly appreciate the time and effort they invested in engaging in one or more interview rounds, without which this work would not have been possible. We are also grateful to the two energy experts for completing the questionnaires and for their valuable discussions and insights. Finally, we would like to express our special thanks to Katharina Esser for developing the RUB energy system model and implementing the MGA approach in previous work. 

JF and VB acknowledge financial support from the German Federal Ministry for Economic Affairs and Energy through the project ``OPTIMA -- Open Toolbox for Modeling and Integration of dynamic-recursive MGA in Energy System Models''  under research grant number 03EI1121B.

\paragraph{Declaration of interest}
None.

\paragraph{Availability of code and data}
The energy system modelling framework Backbone and the implementation of modelling to generate alternatives are available at \url{https://gitlab.vtt.fi/backbone/backbone}.
A basic version of the RUB's energy system model is available at \url{https://doi.org/10.5281/zenodo.12754354}.
The multi-criteria decision analysis tool ValueDecisions is available at \url{https://eawag.shinyapps.io/ValueDecisions/}.
Further data used or generated for this study is available upon request.

\paragraph{Declaration of generative AI and AI-assisted technologies in the manuscript preparation process}
During the preparation of this work, the authors used ChatGPT for language editing of selected manuscript parts to improve readability and clarity, and assistance with programming tasks.
After using this tool/service, the authors reviewed and edited the content as needed and take full responsibility for the content of the published article.

\footnotesize
\singlespacing
\setlength{\bibsep}{0.5pt}

\bibliographystyle{model5-names.bst}
\biboptions{authoryear}
\bibliography{bibliography.bib}


\appendix

\addcontentsline{toc}{chapter}{Appendix}

\section{Energy system model specifications}
\label{appendix: ESM specifications}

\begin{figure}[H]
\centering
\includegraphics[width = 1\textwidth, trim= {0cm 11cm 9cm 6cm}, clip]{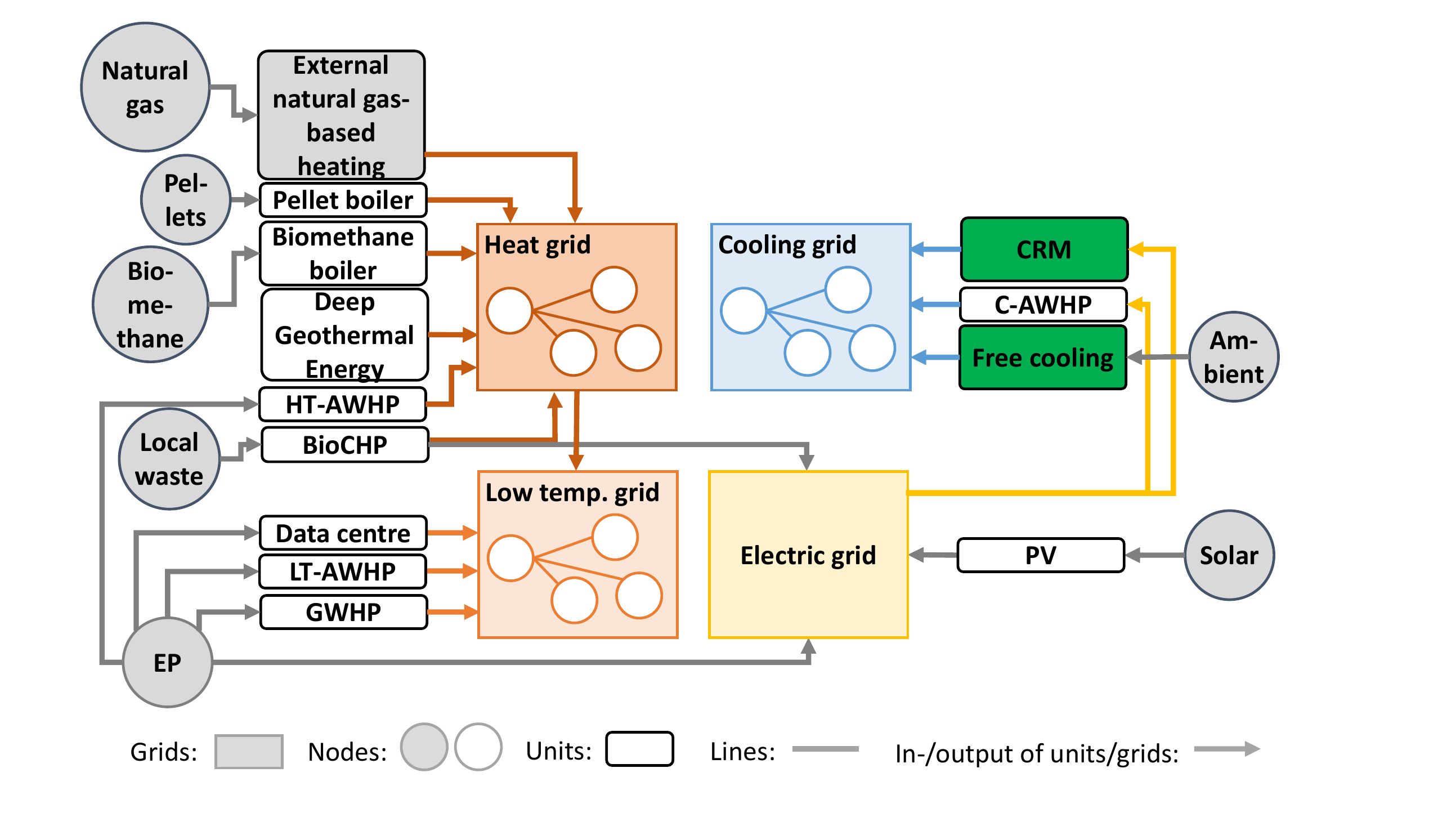}   
\caption{Structure of the energy model of RUB in Backbone for 2045 adapted from \citep{Esser.2024}. Nodes represent energy sources (in grey) or sinks, i.e. individual buildings with energy consumption (in white). Grids group nodes that share a specific form of energy, e.g. heating, cooling or electricity. Lines allow for the transfer of energy between nodes in the same grid. The heating and cooling grids represent physical grids, while the electricity grid refers only to a grid entity in Backbone. Units convert, consume, or generate energy. Units coloured in green are already existing generators, the others are investment options. The external natural gas-based heating supply unit coloured in grey is not available in 2045 due to its emissions.
} 
\label{fig:RUB_GRID}
\end{figure}

\clearpage



\end{document}